\documentstyle[12pt]{article}

\newcommand{\mysection}{\setcounter{equation}{0}\section}

\begin{document}
%-----------------------------
%\format=latex
%\documentstyle[12pt]{article}
%\begin{document}
%--------------------------------------------
%\pagenumbering{}
\vskip 0.2cm
\hfill{YITP-SB-99-65}
\vskip 0.2cm
%\hfill{INLO-PUB-12/99}\\[1cm]
\vskip 0.2cm
\centerline{\large\bf {$\overline{\rm MS}$ Parton densities with NNLO}}
\centerline{\large\bf {heavy flavor matching conditions}}
\vskip 0.2cm
%\centerline {\sc A. Chuvakin }
\centerline {A. Chuvakin, J. Smith 
%\footnote{partially supported
%by the National Science Foundation grant PHY-9722101}
}
\centerline{\it C.N. Yang Institute for Theoretical Physics,}
\centerline{\it State University of New York at Stony Brook,
New York 11794-3840.}
\vskip 0.2cm
\centerline{November 1999}
\vskip 0.2cm
\centerline{\bf Abstract}
\vskip 0.3cm
A study is made of charm and bottom flavor matching conditions 
on parton densities. Starting from an $\overline{\rm MS}$
three-flavor density set, 
where the scale $\mu < m_c$, and using the recently derived 
two-loop matching conditions, we provide a new set of 
four-flavor parton densities where $ m_c \le \mu < m_b$ 
and five-flavor densities where $\mu \ge m_b$.
The effect of the next-to-next-to-leading order matching conditions on the
evolution equations is important for scales just above the transition regions.
This includes the small $x$ and small $Q^2$ domain studied by the H1 and 
ZEUS experiments at HERA. 
At small $x$ the effects of the matching conditions 
never die away even for large $\mu^2$.


\vfill
%\end{document}

%\format=latex
%\documentstyle[12pt]{article}
%\pagestyle{myheadings}  
%\begin{document}
%------------------This is Section 1---------------------------------
\mysection{Introduction}
%----------------------------------------------------------
%\topmargin=0in
%\headheight=0in
%\headsep=0in
%\oddsidemargin=7.2pt
%\evensidemargin=7.2pt
%\footheight=1in
%\marginparwidth=0in
%\marginparsep=0in
%\textheight=9in
%\textwidth=6in
\newcommand{\be}{\begin{eqnarray}}
\newcommand{\ee}{\end{eqnarray}}

Quantum chromodynamic predictions for experimental cross sections and
distributions in perturbation theory rely heavily on accurate
knowledge of parton densities. Several groups \cite{cteq5},
\cite{grv98}, \cite{mrst98} have 
extracted these densities from global fits to data with the
latest theoretical information on $\overline{\rm MS}$ coefficient functions.
At present the evolution of these densities via the  
Altarelli-Parisi (AP) equations
\cite{ap} uses the information on leading order (LO) 
and next-to-leading order (NLO) \cite{cfr} splitting functions. Unfortunately  
the next-to-next-to-leading order (NNLO)
splitting functions are not known but some interesting pieces of information
are available \cite{mom}, \cite{nevo}, \cite{grac}. 
The complete splitting functions should be known soon.

The description of heavy quarks within this analysis recently
received a lot of attention due to the data on deep
inelastic production of $D^*$ mesons from HERA \cite{H1}, \cite{ZEUS}.
The global fitting groups have adopted different approaches.
The CTEQ5 analysis describes charm and bottom densities via
the so-called ACOT prescription \cite{acot}, which is a one-loop matching 
condition between three-flavor and four-flavor densities at the 
scale $\mu = m_c$ and a
corresponding one-loop matching between the four-flavor and five-flavor
densities at the scale $\mu = m_b$. 
This is not done in the MRST density sets. Instead they impose 
matching conditions that the logarithmic derivative
of the deep inelastic structure functions 
with regard to the scale $\mu$ should be continuous, see \cite{thro}
for details. This yields different charm and bottom densities.
The GRV group \cite{grs} adopt the
approach that one does not need any densities other than a three-flavor
$\overline{\rm MS}$ set because the convolution of these densities
with the NLO heavy quark coefficient functions provided 
in \cite{lrsn} yields an excellent fit to the presently available data 
on $F_{2,c}(x,Q^2,m_c^2)$, which is stable under scale variations. By
never taking the limit that $m_c \rightarrow  0$ the theoretical prediction
has no collinear singularity problem. Note that the physical threshold
for a heavy quark antiquark pair is at $Q^2(1-x)/x = 4 m^2$. However
the physical threshold is distinct from the matching scale where one
switches between parton density sets.

The parton densities with $n_f$ and $n_f + 1$ flavors are
related by a set of operator matrix elements (OME's).
The order $\alpha_s^2$ OME's were recently derived in \cite{bmsn1}. 
They contain terms
with $\ln^i(\mu^2/m^2)$ $i=1,2$ as well as non-logarithmic terms. 
They have the property that the $n_f + 1$ flavor densities
vanish in LO and in NLO when the scale $\mu = m$. In NNLO there
are finite $x$-dependent discontinuities at this scale,     
which we refer to as NNLO matching conditions. In this respect the
matching conditions on the parton densities are similar to those for 
the two-loop running coupling constant derived in \cite{2loop}, \cite{2loop2}.
These NNLO matching conditions can be important numerically. 
Note that the running coupling constant
is not small at the scale $\mu = m_c$. The charm density constructed 
from the two-loop matching conditions has been used 
in a recent study of variable flavor number schemes for the 
charm component of the deep inelastic structure
functions \cite{csn}.
Here we would like to present a complete set of parton densities
for light (u,d,s,g) and heavy (c,b) partons
which satisfy the NNLO boundary conditions and are evolved with
NLO splitting functions. 
When the three-loop anomalous dimensions become
available they can be included in this analysis. At present
our study primarily focuses on the effects of the 
discontinuous matching conditions and the differences between 
the resulting three, four and five-flavor parton
densities as the scale $\mu$ increases through the regions
$ m_c \le \mu < m_b$ and $\mu \ge m_b$ respectively.

In Sec. II we present some technical details about
our choice of input parton densities and the two-loop matching conditions. 
All densities are derived in the $\overline{\rm MS}$ scheme. 
Then in Sec.III we give plots of the parton densities. 
Finally comparisons are made
between the presently available NLO sets and our (NNLO) set. 
A discussion of the solutions of the evolution equations
is given in the Appendix. 
We intend to make our computer package for these densities available
in due course.  

%\end{document}

%\documentstyle[12pt]{article}
%\begin{document}

\mysection{The parton densities}

We present a consistent set of $\overline{\rm MS}$ parton 
densities containing three, four and five flavors for scales satisfying 
$\mu < m_c$, $m_c \le \mu < m_b$ and $\mu \ge m_b$ respectively.
The evolution of the densities is done with our own computer
code written in C++ and some details are given in the Appendix.
The code uses the direct $x$-space method 
to solve the evolution equation \cite{ap} similar to
that in \cite{botje} and allows us to evolve both light and heavy 
parton densities in
LO, NLO and NNLO (the latter using the NLO weights). The weights for the
calculation are computed analytically from the
LO and NLO \cite{cfr} $\overline{\rm MS}$
splitting functions thus removing possible
instabilities in the numerical integrations. 
Hence the program is very efficient and fast.
The results from the evolution code have been 
thoroughly checked against the tables in the HERA report \cite{brnv}.
We use weights in LO and NLO for $n_f=3,4,5$ for
evolving the gluon $f_g^{\rm S}(n_f,x,\mu^2)$, the singlet quark densities
$f_q^{\rm S}(n_f,x,\mu^2)$, the
non-singlet valence quark densities $f_{k-\bar k}(n_f,x,\mu^2)$
and the non-singlet sea quark densities
$f_q^{\rm NS}(n_f,x,\mu^2)$.
As the scale increases across the charm and bottom matching points 
the sets are redefined to include densities
for the $c$ and $b$ heavy quarks. 
The program allows us to use LO, NLO or NNLO matching conditions 
for the generation of the heavy flavor densities.

The number densities are defined as
%(2.1)
\begin{eqnarray}
\label{eqn2.1}
&&  f_{k-\bar k}(n_f,x,\mu^2) \equiv f_k(n_f,x,\mu^2) - f_{\bar k}(n_f,x,\mu^2) 
\, , \, k=u,d
\nonumber\\[2ex]
&& f_{k+\bar k}(n_f,x,\mu^2) \equiv f_k(n_f,x,\mu^2) + f_{\bar k}(n_f,x,\mu^2) 
\, ,\,k=u,d,s,c,b
\nonumber\\[2ex]
\end{eqnarray}
% (2.3)
\begin{equation}
\label{eqn2.2}
 f_q^{\rm S}(n_f,x,\mu^2) = \sum_{k=1}^{n_f} f_{k+\bar k}(n_f,x,\mu^2) 
%\nonumber\\[2ex]
\end{equation}
% (2.3)
\begin{equation}
\label{eqn2.3}
 f_q^{\rm NS}(n_f,x,\mu^2) = f_{k+\bar k}(n_f,x,\mu^2) 
-\frac{1}{n_f} f_q^{\rm S}(n_f,x,\mu^2) \,.
\end{equation}

We start our LO evolution using the following input 
from \cite{grv98} at $\mu_0^2=\mu_{\rm LO}^2=0.26$ $({\rm GeV/c}^2)^2$ 
% (A.4)
\begin{eqnarray}
\label{eqn2.4} 
x f_{u-\bar u}(3,x,\mu_0^2)& = &
xu_v(x,\mu_{\rm LO}^2) \nonumber\\
& = & 1.239\,\,x^{0.48}\,(1-x)^{2.72}\, 
(1-1.8\sqrt{x} + 9.5x) \nonumber\\
x f_{d-\bar d}(3,x,\mu_0^2) & = &    
xd_v(x,\mu_{\rm LO}^2) \nonumber\\
& = & 0.614\,\, (1-x)^{0.9}\,\, 
xu_v(x,\mu_{\rm LO}^2)\nonumber\\
x (f_{\bar d}(3,x,\mu_0^2)-f_{\bar u}(3,x,\mu_0^2)) & = &
x\Delta(x,\mu_{\rm LO}^2) \nonumber \\
& = & 0.23 \,\, x^{0.48}\,(1-x)^{11.3}\,
(1-12.0\sqrt{x} + 50.9x)\nonumber\\
x (f_{\bar d}(3,x,\mu_0^2)+f_{\bar u}(3,x,\mu_0^2)) & = &
x(\bar{u}+\bar{d})(x,\mu_{\rm LO}^2) \nonumber \\
& = & 1.52\,\, x^{0.15}\, (1-x)^{9.1}\,
(1-3.6\sqrt{x} + 7.8x)\nonumber\\
x f_g(3,x,\mu_0^2) & = & 
xg(x,\mu_{\rm LO}^2) \nonumber \\
 & = & 17.47\,\, x^{1.6}\, (1-x)^{3.8}\nonumber\\
x f_{s}(3,x,\mu_0^2) =   
x f_{\bar s}(3,x,\mu_0^2)  & = &  xs(x,\mu_{\rm LO}^2) \nonumber \\  
& = & x\bar{s}(x,\mu_{\rm LO}^2) = 0 \,.
%\nonumber\\
\end{eqnarray}
Here $\Delta\equiv\bar{d}-\bar{u}$ is used to construct 
the non-singlet combination.
We start the corresponding NLO evolution using the following input 
from \cite{grv98} at $\mu_0^2=\mu_{\rm NLO}^2=0.40$ $({\rm GeV/c}^2)^2$ 
% (A.5)
\begin{eqnarray}
\label{eqn2.5}
x f_{u-\bar u}(3,x,\mu_0^2) & = & 
xu_v(x,\mu_{\rm NLO}^2) \nonumber \\
& = & 0.632\,\,x^{0.43}\,(1-x)^{3.09}\, 
(1+18.2x)\nonumber\\
x f_{d-\bar d}(3,x,\mu_0^2) & = &    
xd_v(x,\mu_{\rm NLO}^2) \nonumber \\
& = & 0.624\,\, (1-x)^{1.0}\,\, 
xu_v(x,\mu_{\rm NLO}^2)\nonumber\\
x (f_{\bar d}(3,x,\mu_0^2)-f_{\bar u}(3,x,\mu_0^2)) & = &
x\Delta(x,\mu_{\rm NLO}^2) \nonumber \\
& = & 0.20 \,\, x^{0.43}\,(1-x)^{12.4}\,
(1-13.3\sqrt{x} + 60.0x)\nonumber\\
x (f_{\bar d}(3,x,\mu_0^2)+f_{\bar u}(3,x,\mu_0^2)) & = &
x(\bar{u}+\bar{d})(x,\mu_{\rm NLO}^2) \nonumber \\
& = & 1.24\,\, x^{0.20}\, (1-x)^{8.5}\,
(1-2.3\sqrt{x} + 5.7x)\nonumber\\
x f_g(3,x,\mu_0^2) & = & xg(x,\mu_{\rm NLO}^2) \nonumber \\ 
& = & 20.80\,\, x^{1.6}\, (1-x)^{4.1}\nonumber\\
x f_{s}(3,x,\mu_0^2) =   
x f_{\bar s}(3,x,\mu_0^2) & = &
xs(x,\mu_{\rm NLO}^2)\nonumber \\
 & = & x\bar{s}(x,\mu_{\rm NLO}^2) = 0.
\end{eqnarray}
From the above densities we form the combinations that we
evolve and step across thresholds. The NNLO densities for 
$\mu_{\rm NLO}^2 < \mu^2 < m_{\rm c}^2$ are replaced with NLO densities. 
The heavy quark masses $m_c = 1.4$ ${\rm GeV/c}^2$, $m_b = 4.5$
$ {\rm GeV/c}^2$
are used throughout the calculation together with the exact expression 
for the running coupling constant $\alpha_s(\mu^2)$,
represented as the solution of the following differential equation
%(2.6)
\begin{equation}
\label{eqn2.6}
\frac{d\, \alpha_s(\mu^2)}{d\, \ln(\mu^2)}= 
-\frac{\beta_0}{4\pi}\, \alpha_s^2(\mu^2)
\, -\, \frac{\beta_1}{16\pi^2}\, \alpha_s^3(\mu^2)
\end{equation}
or in the implicit form
%(2.7)
\begin{equation}
\label{eqn2.7}
\ln\frac{\mu^2}{(\tilde{\Lambda }^{(n_f)}_{\rm EXACT})^2} 
= \frac{4\pi} {\beta_0\alpha_s(\mu^2)}\,-\, \frac{\beta_1}{\beta_0^2} \, \ln
\left[ \frac{4\pi}{\beta_0\alpha_s(\mu^2)}\, +\, \frac{\beta_1}{\beta_0^2}
\right] ,
\end{equation}
where $\beta_0=11 - 2n_f/3$ and $\beta_1=102 - 38n_f/3$.
The values for $\tilde{\Lambda}^{(n_f)}_{\rm EXACT}$
are carefully chosen to obtain accurate matching at the scales
$m_c^2$ and $m_b^2$ respectively. We used the values 
$\tilde {\Lambda}^{(3,4,5,6)}_{\rm EXACT}=299.4,$\,
$246,\, 167.7,$ $\, 67.8$ 
MeV$/c^2$
respectively in the exact formula 
(which yields 
$\alpha_s^{\rm EXACT}(m_Z^2)=0.114$,
$\alpha_s^{\rm EXACT}(m_b^2)=0.205$,
$\alpha_s^{\rm EXACT}(m_c^2)=0.319$,
$\alpha_s^{\rm EXACT}(\mu_{\rm NLO}^2)=0.578$
) and
$\Lambda_{\rm LO}^{(3,4,5,6)} = 204,$\,
$175,\, 132,$\,$ 66.5$ MeV$/c^2$ 
respectively 
(which yields 
$\alpha_s^{\rm LO}(m_Z^2) = 0.125$,
$\alpha_s^{\rm LO}(m_b^2) = 0.232$,
$\alpha_s^{\rm LO}(m_c^2) = 0.362$,
$\alpha_s^{\rm LO}(\mu_{\rm LO}^2) = 0.763$
) 
for the LO formula. Note that we have not used the two-loop matching
of the running coupling constant $\alpha_s(\mu^2)$
at the same scales from \cite{2loop},\cite{2loop2} to focus on the
matching conditions on the flavor densities. 
Numerically the discontinuity in the running coupling
constant across the charm threshold is aproximately two parts in one
thousand, which is far too small to affect our results.

Three flavor evolution proceeds from the initial $\mu_0^2$ to the 
scale $\mu^2=m_c^2=1.96$ $({\rm GeV/c}^2)^2$. 
At this point the charm density is then defined by 
\begin{eqnarray}
\label{eqn2.8}
f_{c + {\bar c}}(n_f+1,m_c^2)&=&
a_s^2 (n_f,m_c^2)\Big [\tilde A_{Qq}^{\rm PS}
(1)\otimes f_q^{\rm S}(n_f, m_c^2)
\nonumber\\[2ex]  
&& + \tilde A_{Qg}^{\rm S}(1) 
\otimes f_g^{\rm S}(n_f, m_c^2)\Big ] \,,
\end{eqnarray}
with $n_f=3$ and $a_s = \alpha_s/4\pi$.
We have suppressed the $x$ dependence to make the notation more compact.
The $\otimes$ symbol denotes the convolution integral
$f\otimes g=\int f(x/y)g(y)dy/y$, where $x \le y \le 1$ 
The OME's $\tilde A^{\rm PS}_{Qq}(\mu^2/m_c^2)$,
$\tilde A^{\rm S}_{Qg}(\mu^2/m_c^2)$
are given in \cite{bmsn1}. 
The reason for choosing the matching scale $\mu$ 
at the mass of the charm quark $m_c$ is that all the $\ln(\mu^2/m_c^2)$ 
terms in the OME's vanish at this point leaving only the
nonlogarithmic pieces in the order $\alpha_s^2$ OME's to contribute to
the right-hand-side of Eq.(2.8). Hence the LO and NLO charm densities
vanish at the scale $\mu=m_c$. The NNLO charm density starts off with 
a finite $x$-dependent shape in order $a_s^2$. Note that we then order 
the terms on the right-hand-side of Eq.(2.8) so that the result contains 
a product of NLO OME's and LO parton densities. The result is then of 
order $a_s^2$ and should be multiplied by order $a_s^0$ coefficient 
functions when forming the deep inelastic structure functions. 

The four-flavor gluon density is also generated at the
matching point in the same way. At $\mu = m_c$ we define
%(2.9)
\begin{eqnarray}
\label{eqn2.9}
f_g^{\rm S}(n_f+1, m_c^2) &=& f_g^{\rm S}(n_f, m_c^2)
\nonumber\\%[2ex]
&&  + a_s^2(n_f,m_c^2) \Big [ A_{gq,Q}^{\rm S}(1)
\otimes f_q^{\rm S}(n_f, m_c^2) \,, 
\nonumber\\%[2ex]  
&&+ A_{gg,Q}^{\rm S}(1)
\otimes f_g^{\rm S}(n_f, m_c^2) \Big ]\,.
\end{eqnarray}
The OME's $A^{\rm S}_{gq,Q}(\mu^2/m_c^2)$, $A^{\rm S}_{gg,Q}(\mu^2/m_c^2)$
are given in \cite{bmsn1}. 
The four-flavor light quark (u,d,s) densities 
are generated using
\begin{eqnarray}
\label{eqn2.10}
f_{k+\bar k}^{\rm}(n_f+1,m_c^2) &=& f_{k+\bar k}(n_f,m_c^2) 
\nonumber\\%[2ex]
&& + a_s^2(n_f,m_c^2) A_{qq,Q}^{\rm NS}(1)
\otimes f_{k+\bar k}(n_f, m_c^2)\,.
%\nonumber\\%[2ex]
\end{eqnarray}
The OME $A^{\rm NS}_{qq,Q}(\mu^2/m_c^2)$ is given in \cite{bmsn1} 
and the {\it total} four-flavor singlet quark density in Eq.(2.2) 
follows from the sum of Eqs.(2.8) and (2.10).
The nonsinglet density then follows from Eq.(2.1).
In Eqs.(2.9) and (2.10) $n_f = 3$. The remarks after Eq.(2.8) are relevant 
here too.

Next the resulting four-flavor densities are evolved using the 
four-flavor weights
in either LO or NLO up to the scale
$\mu^2=20.25$ $({\rm GeV/c}^2)^2$. 
The bottom quark density is then generated at this point using 
\begin{eqnarray}
\label{eqn2.11}
f_{b + {\bar b}}(n_f+1,m_b^2)&=&
a_s^2 (n_f,m_b^2)\Big [\tilde A_{Qq}^{\rm PS}
(1)\otimes f_q^{\rm S}(n_f, m_b^2)
\nonumber\\%[2ex]  
&&+ \tilde A_{Qg}^{(\rm S)}(1) 
\otimes f_g^{\rm S}(n_f, m_b^2)\Big ] \,,
\end{eqnarray}
and the gluon and light quark densities (which now include charm) 
are generated using
Eqs.(\ref{eqn2.8})-(\ref{eqn2.10}) 
with $n_f =4$ and replacing $m_c^2 $ by $m_b^2$. 
Therefore only the nonlogarithmic terms in the order $a_s^2$ OME's 
contribute to the matching conditions on the bottom quark density.
Then all the densities are evolved up to higher $\mu^2$ as a five-flavor set
with either LO or NLO splitting functions. This is valid until
$\mu = m_t \approx 175$ ${\rm GeV/c}^2$ above which one should switch
to a six-flavor set. We do not implement this step because the top
quark density would be extremely small.

The procedure outlined above generates a full set of parton densities
(gluon, singlet, non-singlet light and heavy quark densities,) 
for any $x$ and $\mu^2$ from the three-flavor LO and NLO inputs 
in Eqs.(\ref{eqn2.4}) and (\ref{eqn2.5}).
Note that one could also use the formulae above in fixed order perturbation 
theory. In this case the four-flavour densities are
defined by extending the integrals on the right-hand sides of 
Eqs.(2.8)-(2.11) to 
%(2.12)
\begin{eqnarray}
\label{eqn2.12}
f_{c + {\bar c}}(n_f+1,\mu^2)&=&
a_s(n_f,\mu^2) \tilde A_{Qg}^{\rm S}\Big(\frac{\mu^2}{m_c^2}\Big) \otimes
f_g^{\rm S}(n_f, \mu^2) 
\nonumber\\%[2ex]  
&& + a_s^2 (n_f,\mu^2)\Big [\tilde A_{Qq}^{\rm PS}
\Big(\frac{\mu^2}{m_c^2}\Big)\otimes f_q^{\rm S}(n_f, \mu^2)
\nonumber\\[2ex]  
&& + \tilde A_{Qg}^{\rm S}\Big(\frac{\mu^2}{m_c^2}\Big) 
\otimes f_g^{\rm S}(n_f, \mu^2)\Big ] \,,
\end{eqnarray}
% (2.13)
\begin{eqnarray}
\label{eqn2.13}
f_g^{\rm S}(n_f+1, \mu^2) &=& f_g^{\rm S}(n_f, \mu^2)
\nonumber\\%[2ex]
&& +a_s(n_f,\mu^2)A_{gg,Q}^{\rm S}(\frac{\mu^2}{m_c^2}) \otimes 
f_g^{\rm S}(n_f, \mu^2)
\nonumber\\%[2ex]  
&& + a_s^2(n_f,\mu^2) \Big [ A_{gq,Q}^{\rm S}(\frac{\mu^2}{m_c^2})
\otimes f_q^{\rm S}(n_f, \mu^2) \,, 
\nonumber\\%[2ex]  
&&+ A_{gg,Q}^{\rm S}(\frac{\mu^2}{m_c^2})
\otimes f_g^{\rm S}(n_f, \mu^2) \Big ]\,,
\end{eqnarray}
and
%(2.14)
\begin{eqnarray}
\label{eqn2.14}
f_{k+\bar k}^{\rm}(n_f+1,\mu^2) &=& f_{k+\bar k}(n_f,\mu^2) 
\nonumber\\%[2ex]
&& + a_s^2(n_f,\mu^2) A_{qq,Q}^{\rm NS}\Big(\frac{\mu^2}{m_c^2}\Big)
\otimes f_{k+\bar k}(n_f, \mu^2)\,,
\nonumber\\%[2ex]
\end{eqnarray}
%(2.15)
for $n_f = 3$ and $m_c^2 \le \mu^2 <  m_b^2$. 
Then the five-flavor densities are defined by 
\begin{eqnarray}
\label{eqn2.15}
f_{b + {\bar b}}(n_f+1,\mu^2)&=&
a_s(n_f,\mu^2) \tilde A_{Qg}^{\rm S}\Big(\frac{\mu^2}{m_b^2}\Big) \otimes
f_g^{\rm S}(n_f, \mu^2) 
\nonumber\\[2ex]
&& + a_s^2 (n_f,\mu^2)\Big [\tilde A_{Qq}^{\rm PS}
\Big(\frac{\mu^2}{m_b^2}\Big)\otimes f_q^{\rm S}(n_f, \mu^2)
\nonumber\\%[2ex]  
&&+ \tilde A_{Qg}^{\rm S)}\Big(\frac{\mu^2}{m_b^2}\Big) 
\otimes f_g^{\rm S}(n_f, \mu^2)\Big ] \,,
\end{eqnarray}
for $n_f=4$ and $\mu^2 \ge m_b^2$. Also one should
replace $n_f=4$ and $m_c^2$ by $m_b^2$ in Eqs.(2.12)-(2.14). 
In this case no four-flavor or five-flavour evolution
is made so the logarithmic terms in $\mu^2/m_c^2$ and/or $\mu^2/m_b^2$ 
are not summed. We will show the differences between this fixed order
perturbation theory (FOPT) treatment and the evolved treatment in the
next Section.

%\end{document}

%%%%%%%%%%%%%%%%%%%%%%%%%%%%%%%%%%%%%%%%%%%%%%%%%%%%%%%%%%%%%%%%%%%%%%%
%\format=latex
%\documentstyle[12pt]{article}
%\pagestyle{myheadings}  
%\begin{document}
%------------------This is Section 3---------------------------------
\mysection{Results}
%----------------------------------------------------------
Here we present results from the evolution 
of the parton densities. The inputs are the three-flavor 
densities at $\mu_0$ in Eqs.(2.4),(2.5) which are evolved 
up to the scale $\mu = m_c = 1.4$ ${\rm GeV/c}^2$. During this
evolution the number of light flavors $n_f = 3$ 
in both the $\overline{\rm MS}$ splitting functions and 
the running coupling constant. 

We start by giving the four-flavor densities, where $n_f=4$,
in the region between 
$m_c \le \mu < m_b$ which follow by evolution
from the matching conditions in Eqs. (2.8)-(2.10). 
We present results at the 
scales $\mu^2 =$ 1.96, 2, 3, 4, 5, 10 and 20 
in units of $({\rm GeV/c}^2)^2$. 
First we show the charm density $xc^{\rm NNLO}(4,x,\mu^2)$  
in two ranges (a) $10^{-5} < x < 1$  and (b) $10^{-2} < x < 1$
in Figs.1(a) and 1(b) respectively. 
We notice that this density starts off
negative at small $x$ but it is positive at large $x$
so that the momentum sum rule is satisfied.
To show the effect of resumming the logarithmic terms via the
evolution equation from the charm threshold
as compared with just computing the
integrals in Eq. (2.12) at all scales $\mu$ we show in Fig.1(c) the ratios 
$R^{\rm NNLO}_{c}(x,\mu^2)
=xc_{\rm EVOLVED}(4,x,\mu^2)$/$xc_{\rm FOPT}(4,x,\mu^2)$. 
Here FOPT stands for fixed order perturbation
theory. The effects of the evolution are especially significant
at small $x$ and large $x$. Notice that the discontinuity at
$x \approx 0.01$ is caused by the change in sign of the NNLO 
charm quark density. 
For a comparison we have also shown the NLO
results from the MRST98 set 1 \cite{mrst98} and CTEQ5HQ \cite{cteq5}
parton density sets in Figs. 1(d) and (e). These groups
use different input densities so a direct comparison does not have
any true significance. Nevertheless our density is larger than 
the MRST98 set 1 result and smaller than the CTEQ5HQ result at 
small $x$ and large $\mu^2$. 

In Fig.2(a) we show the four-flavor gluon density 
$xg^{\rm NNLO}(4,x,\mu^2)$ in the same range
$10^{-5} < x < 1$ for the same scales as in Fig.1. 
We also show in Fig.2(b) the ratios 
$R^{\rm NNLO}_{g}(x,\mu^2)
=xg_{\rm EVOLVED}(4,x,\mu^2)$/$xg_{\rm FOPT}(4,x,\mu^2)$ 
for the same scales, where we use Eq.(2.13) for the FOPT density.
The effect of the suppression of the charm
density at small $x$ translates into an increase of the gluon density
at small $x$. For comparison we show the three-flavor NLO gluon 
density in Fig. 2(c). We have also shown the NLO
results from the MRST98 set 1 and CTEQ5HQ 
parton density sets in Figs. 2(d) and (e). These densities do not
increase as rapidly at large $\mu^2$ because they use different
inputs.  

In Fig.3(a) we show the singlet quark 
density $x\Sigma^{\rm NNLO}(4,x,\mu^2)$ 
in the range $10^{-5} < x < 1$ for the same scales as above.
Then is Fig.3(b) we show the ratios 
$R^{\rm NNLO}_{\Sigma}(x,\mu^2)=x\Sigma_{\rm EVOLVED}(4,x,\mu^2)$
/$x\Sigma_{\rm FOPT}(4,x,\mu^2)$, where we use Eq.(2.14) for the FOPT
density.
This ratio shows increases or decreases depending on the $x$ and
$\mu^2$ values.
It is appreciably smaller at large $\mu^2$, which reflects the 
differences between the three-flavor and four-flavor gluon densities.
For comparison we show the three-flavor NLO density in Fig.3(c).

Next in Fig.4(a) we show the nonsinglet density 
$x\sigma^{\rm NNLO}(4,x,\mu^2)$, where $\sigma =(u+\bar u)/2$, in the range
$10^{-5} < x < 1$ for the same scales. 
In Fig.4(b) we show the ratios 
$R^{\rm NNLO}_{\sigma}(x,\mu^2)
=x\sigma_{\rm EVOLVED}(4,x,\mu^2)$/$x\sigma_{\rm FOPT}(4,x,\mu^2)$ 
for the same scales, where we use Eq.(2.14) for the FOPT
density. The ratio is significantly below unity at large $x$. 
In Fig.4(c) we show the three-flavor NLO result. The difference is small.

We complete our presentation of four-flavor densities by showing
in Fig.5(a) the strange quark density $xs^{\rm NNLO}(4,x,\mu^2)$ in the range
$10^{-5} < x < 1$. In Fig.5(b) we show the ratios 
$R^{\rm NNLO}_{g}(x,\mu^2)
=xs^{\rm EVOLVED}(4,x,\mu^2)$/$xs^{\rm FOPT}(4,x,\mu^2)$ 
for the same scales.
In Fig.5(c) we show the three-flavor NLO density, where again 
the difference is small. 
We have checked that these densities satisfy the momentum sum rule for
four flavors. 

Now we move up in scale to consider $\mu > m_b = 4.5$ ${\rm GeV/c}^2$
which is the five-flavor region. The parton densities in this region
are now generated from the previous four-flavor set by 
using the conditions in the Eqs.(2.8)
-(2.10) with $n_f = 4$ at $\mu^2 = m_b^2$ and replacing  
$m_c^2$ by $m_b^2$. Here we show plots for the scales
$\mu^2 =$20.25, 25, 30, 40, and 100 in units of 
$({\rm GeV/c}^2)^2$.  

The first density to consider is the bottom quark density.
We show in Fig.6(a) $xb^{\rm NNLO}(5,x,\mu^2)$ 
in the range $10^{-5} < x < 1$ for the scales mentioned above.
Notice that it is negative for small $x$ and small $\mu^2$,
The region $10^{-2} < x < 1$ is shown in Fig. 6(b) 
to demonstrate that the density is positive for large $x$.
In this respect it is like the charm density in the region
just above the four-flavor matching point.
In Fig.6(c) we show the ratios 
$R^{\rm NNLO}_{g}(x,\mu^2)
=xb^{\rm EVOLVED}(5,x,\mu^2)$/$xb^{\rm FOPT}(5,x,\mu^2)$,
where Eq.(2.15) is used for the FOPT density.
Here it is clear that the effect of the evolution is 
appreciable at small $x$. 
For a comparison we have also shown the NLO
results from the MRST98 set 1 and CTEQ5HQ 
parton density sets in Figs. 1(d) and (e). Note that these groups
use different input densities so a direct comparison does not have
any true significance. Nevertheless both bottom densities are 
larger than ours at small $x$ and large $\mu^2$.

The five-flavor charm density $xc^{\rm NNLO}(5,x,\mu^2)$ 
is shown in Fig.7(a) in the range $10^{-5} < x < 1$ for 
the same scales.
In this case the ratios of the evolved densities to the
FOPT densities is very close to unity for all $x$ and
$\mu^2$ values so we do not show a plot.
For a comparison we have also shown the NLO
results from the MRST98 set 1 and CTEQ5HQ
parton density sets in Figs. 1(b) and (c). The former has a smaller density
at large $\mu^2$ and the latter a larger density. 

Next we show in Fig.8(a) the gluon density 
$xg^{\rm NNLO}(5,x,\mu^2)$ in the range $10^{-5} < x < 1$ for 
the same scales.
In this case the ratios of the evolved densities to the
FOPT densities is very close to unity for all $x$ and
$\mu^2$ values so we do not show a plot.
In Fig.8(b) we show the corresponding three-flavor NLO density.
The latter is larger at small $x$ and large $\mu^2$.

In Fig.9(a) we show the nonsinglet quark
density $x\Sigma^{\rm NNLO}(5,x,\mu^2)$, where $\sigma = (u + \bar u)/2$,
in the range $10^{-5} < x < 1$ for the same scales.
Also in this case the ratios of the evolved densities to the
FOPT densities is very close to unity for all $x$ and
$\mu^2$ values so we do not show any plot.
In Fig.9(b) we show the corresponding three-flavor NLO density.

In Fig.10(a) we show the nonsinglet quark
density $x\sigma^{\rm NNLO}(5,x,\mu^2)$ 
in the range $10^{-5} < x < 1$ for the same scales.
Also in this case the ratios of the evolved densities to the
FOPT densities is very close to unity for all $x$ and
$\mu^2$ values so we do not show any plot.
In Fig.10(b) we show the corresponding three-flavor NLO density.

In Fig.11(a) we show the strange quark density $xs^{\rm NNLO}(5,x,\mu^2)$ 
in the range $10^{-5} < x < 1$ for the same scales. 
Also in this case the ratios of the evolved densities to the
FOPT densities is very close to unity for all $x$ and
$\mu^2$ values so we do not show any plot.
In Fig.11(b) we show the corresponding three-flavor NLO density.
We have checked that the five-flavour densities satisfy the momentum
sum rule. 

The above plots demonstrate that the NNLO matching conditions
do influence the parton densities appreciably in regions just above $\mu = m_c$
and $\mu = m_b$. This has consequences for the analysis of HERA
experiments because a lot of the data is at small $x$ and 
small values of the scale $Q^2$. In fact all of the data for 
$x < 10^{-4}$ has $Q^2 < 100$ $({\rm GeV/c}^2)^2$.
Even for the scale $\mu$ much larger than $m_b$ the boundary 
conditions are still important. For $\mu \ge 10$ ${\rm GeV/c}^2$
for example the rapid rise of the 
five-flavour gluon density means that it
dominates over all the other parton densities at small $x$.
There is roughly a ten percent difference between the three-flavor
and five-flavor densities at small $x$ and large $\mu^2$, 
which can be important for precision phenomenology. 

%---------------------------------------------------------
\medskip

ACKNOWLEDGMENTS

This research was partially supported by the National Science 
Foundation grant PHY-9722101.  We thank M. Botje, E. Laenen and 
W.L. van Neerven for very useful discussions.

%\end{document}

%\end{document}

\appendix
%threshaa.tex
%\documentstyle[12pt]{article}
%\begin{document}
%\bigskip
\mysection*{Appendix A}
\setcounter{section}{1}

All the splitting functions in the Altarelli-Parisi (AP) equations
can be expanded as a perturbation series 
in $\alpha_s$ into LO and NLO terms as follows
%(A.1)
\begin{equation}
P=P^{(0)}+\frac{\alpha_s}{2\pi} P^{(1)}.
\end{equation}
The non-singlet combinations of the $q_r(\bar q_r)$ to $q_s(\bar q_s)$
splitting functions, 
where the subscripts $r,s$ denote the flavors of the (anti)quarks and
satisfies $r,s=1,\cdots,n_f$,
can be further decomposed into a flavor diagonal part proportional
to $\delta_{rs}$ and a flavor independent part.
In LO there is only one non-singlet splitting function $P_{qq}$
but in NLO it is convenient to form two combinations
from $P_{qq} $ and $P_{q \bar q}$ as follows
%(A.2)
\begin{eqnarray}
P_+=P_{qq}+P_{q{\bar q}}
\nonumber \\
P_-=P_{qq}-P_{q{\bar q}}.
\end{eqnarray}
These splitting functions are used to evolve two independent types 
of non-singlet densities, which will be called plus and minus respectively. 
They are given by
%(A.3)
\begin{eqnarray}
f^+_i = f_q^{\rm NS}(n_f,x,\mu^2)
\nonumber \\
f^-_j = f_{k-\bar k}(n_f,x,\mu^2).
\end{eqnarray}
The easiest way to explain the indices in these equations is by explicitly
giving the combinations we use. For $j=1,2$
%(A.4)
\begin{equation}
f_1^-  =  u - \bar u\,, \, f_2^-  =  d - \bar d\,,
\end{equation}
which are used for all flavour density sets.
Then for three-flavor densities $i = 1,2,3$ and we define
%(A.5)
\begin{eqnarray}
&& f_1^+  =  u + \bar u - \Sigma(3)/3\,,\quad  \quad
   f_2^+  =  d + \bar d - \Sigma(3)/3\,, \nonumber \\
&& f_3^+  =  s + \bar s - \Sigma(3)/3\,, 
\end{eqnarray}
where $\Sigma(3) = f_q^S(3) = u + \bar u + d + \bar d + s + \bar s$.
These densities should be used for $\mu < m_c$.
For four-flavor densities $i = 1,2,3,4$ and we define
%(A.6)
\begin{eqnarray}
&& f_1^+  =  u + \bar u - \Sigma(4)/4\,, \quad \quad
   f_2^+  =  d + \bar d - \Sigma(4)/4\,, \nonumber \\
&& f_3^+  =  s + \bar s - \Sigma(4)/4\,, \quad \quad 
   f_4^+  =  c + \bar c - \Sigma(4)/4\,, 
\end{eqnarray}
where $\Sigma(4) = f_q^S(4) = c + \bar c + \Sigma(3)$.
These should be used for $m_c \le \mu < m_b$.
For five-flavor densites $i = 1,2,3,4,5$ and we define 
%(A.7)
\begin{eqnarray}
&& f_1^+  =  u + \bar u - \Sigma(5)/5\,, \quad \quad 
   f_2^+  =  d + \bar d - \Sigma(5)/5\,, \nonumber \\
&& f_3^+  =  s + \bar s - \Sigma(5)/5\,, \quad \quad 
   f_4^+  =  c + \bar c - \Sigma(5)/5\,, \nonumber \\
&& f_5^+  =  b + \bar b - \Sigma(5)/5\,, 
\end{eqnarray}
where $\Sigma(5) = f^q_S(5) = b + \bar b + \Sigma(4)$.
These should be used for $\mu \ge m_b$.

If we define $t = \ln(\mu^2/(1 ({\rm GeV/c}^2)^2)$ then  
the AP equations that we need to solve are
%(A.8)
\begin{eqnarray}
\label{AP}
\frac{\partial f^{+}_{i}(x,t)}{\partial t}
&=&\frac{\alpha _{s}(t)}{2\pi }\int_x^1 
\frac{dz}{z}P_{+}(\frac{x}{z})f^{+}_{i}(z,t) \,,
\end{eqnarray}
%
%(A.9)
\begin{eqnarray}
\frac{\partial f^{-}_{j}(x,t)}{\partial t}
&=&\frac{\alpha _{s}(t)}{2\pi }\int_x^1 
\frac{dz}{z}P_{-}(\frac{x}{z})f^{-}_{j}(z,t) \,,
\end{eqnarray}
%
%(A.10)
\begin{eqnarray}
\frac{\partial f_{g}(x,t)}{\partial  t}
&=&\frac{\alpha _{s}(t)}{2\pi }\int_x^1
\frac{dz}{z} \left[ P_{gq}(\frac{x}{z})f^{S}_{q}(z,t)
+P_{gg}(\frac{x}{z})f^{S}_{g}(z,t)\right] \,,
\end{eqnarray}
%
%(A.11)
\begin{eqnarray}
\frac{\partial f^{S}_{q}(x,t)}{\partial  t}
&=&\frac{\alpha _{s}(t)}{2\pi }\int_x^1 
\frac{dz}{z} \left[ P_{qq}(\frac{x}{z})f^{S}_{q}(z,t)
+P_{qg}(\frac{x}{z})f^{S}_{g}(z,t)\right] \,,
\end{eqnarray}
where for $\mu < m_c$ we set $i=1,2,3$, $j=1,2$, 
$f_q^S = \Sigma(3)$ and the gluon is a three-flavor gluon.
When $m_c \le \mu < m_b$, we use    
$i=1,2,3,4$, $j=1,2$, $f_q^S = \Sigma(4)$ and the gluon is a four-flavor
gluon.  Finally when $\mu \ge m_b$, we set
$i=1,2,3,4,5$, $j=1,2$, $f_q^5 = \Sigma(5)$ and the gluon is 
a five-flavor gluon.

The densities should satisfy the momentum conservation sum rule
%(A.12)
\begin{eqnarray}
\label{momsum}
&& \int_0^1 x\left[u(x,\mu^2) + d(x,\mu^2) + \bar{u}(x,\mu^2)
+ \bar{d}(x,\mu^2) + 2 s(x,\mu^2) \right.  
\nonumber \\
&& \left. + 2 c(x,\mu^2)\theta(\mu^2 - m_c^2) 
   + 2 b(x,\mu^2)\theta(\mu^2 - m_b^2) + g(x,\mu^2) \right]
 \, dx = 1 \,.
\nonumber \\
\end{eqnarray}
As the quark constituents carry all the 
charge, isospin, strange, charm and bottom quantum numbers of the nucleon
they should also satisfy the other standard sum rules for the conservation 
of these quantities.

There are several methods to solve these equations. Among
them the most popular are to use Mellin 
moments (used by \cite{riemer}, \cite{pasc}, see full list of 
references in \cite{brnv} and \cite{botje}) and to use
the direct $x$-space solution (as in
\cite{miya}, \cite{botje},\cite{pasc}, see also \cite{brnv} ). 
Also the authors in \cite{cori}
describe a method involving Laguerre polynomials, that dates back to early
paper of \cite{cfr}.

Our choice of direct $x$-space method is justified by the necessity to step
densities across matching points using LO, NLO and NNLO boundary conditions.
The procedure of doing this in the Mellin moment method would 
involve converting densities to and from Mellin moments several times.
Using the direct $x$-space method is much more intuitive and
straightforward.
The main features of this method are linear interpolation over a grid
in $x$ and second-order interpolation over a grid in $t$. 
Let us describe the method in more detail to point out where we 
differ from the work in \cite{botje}.

First we consider the $x$-variable in the evolution. Consider the 
right-hand-side of the evolution equation (\ref{AP}) for non-singlet density
%(A.13)
\begin{equation}
I(x_{0})=\int
\frac{dz}{z} \frac{x_{0}}{z} P\left( \frac{x_{0}}{z}\right) 
q\left(z\right) \,, 
\end{equation}
where $x_0 \le z \le 1$ and
%(A.14)
\begin{equation}
q(x)=xf(x) \,,
\end{equation}
and
%(A.15)
\begin{equation}
x_{0} < x_{1} <...< x_{n}<x_{n+1}\equiv 1 \,,
\end{equation}
with \( q(x_{n+1})=q(1)\equiv 0 \).
Between grid points $x_i$ and $x_{i+1}$, $x$ is chosen so that
%(A.16)
\begin{equation}
\label{inth}
q(x) =(1 - y)q(x_{i})+y q(x_{i+1}) \,,
\end{equation}
with $y = (x-x_{i})/(x_{i+1}-x_{i}) $.
Using this relation we convert the integral into a sum
%(A.17)
\begin{equation}
I(x_{0})=\sum_{i=0}^{n+1} w(x_{i},x_{0})q(x_{i}) \,,
\end{equation}
where the weights are
%(A.18)
\begin{eqnarray}
w(x_{0},x_{0})&=&S_{1}(s_{1},s_{0}) 
\nonumber \\
w(x_{i},x_{0})&=&S_{1}(s_{i+1},s_{i})-S_{2}(s_{i},s_{i-1}) \,,
\end{eqnarray}
where $s_{i}=x_{0}/x_{i}$ and
%(A.18)
\begin{eqnarray}
S_{1}(u,v)&=&\frac{v}{v-u}\int_u^v (z-u)P(z)\frac{dz}{z}
\nonumber \\
S_{2}(u,v)&=&\frac{u}{v-u}\int_u^v (z-v)P(z)\frac{dz}{z} \,.
\end{eqnarray}
We have calculated these integrals analytically and the results are
in the computer program.
This leads to the final formula describing the grid for the $x$ variable.
Note that the weights $w^{(0)}$ and $w^{(1)}$ include LO and NLO splitting
functions respectively. Thus, for the singlet case, we have 
%(A.19)
\begin{eqnarray}
\label{xevol}
\frac{d (x_{0}\Sigma (x_{0}))}{d t}
&=&\frac{\alpha _{s}}{2\pi }\sum \left[ w^{(0)}_{qq}(x_{i},x_{0})
+\frac{\alpha _{s}}{2\pi }w_{qq}^{(1)}(x_{i},x_{0})\right] 
x_{i}\Sigma (x_{i})
\nonumber \\
&& + \left[ w^{(0)}_{qg}(x_{i},x_{0})+\frac{\alpha _{s}}{2\pi }
w_{qg}^{(1)}(x_{i},x_{0})\right] x_{i}g(x_{i}) \,.
\end{eqnarray}

Now consider the variation in the variable $t$. For each $x_i$ we 
pick a grid in $t$ labelled by distinct points $t_j$.  
Then the example the non-singlet equation becomes
%(A.20)
\begin{equation}
\label{qevol}
q^{'}(x_{i},t_{j})
=\frac{\alpha _{s}(t_{j})}{2\pi }
\sum_{k=1}^n \left[ w^{(0)}_{\pm }(x_{k},x_{i})
+\frac{\alpha _{s}(t_{j})}{2\pi }w^{(1)}_{\pm }(x_{k},x_{i})\right] 
q(x_{k},t_{j}) \,,
\end{equation}
where $ q^{'}(x_{i},t_{j})$ denotes the derivative with respect to 
$t$ evaluated at $t = t_j$. 
In compact notation this equation can be rewritten as
%(A.21)
\begin{equation}
\label{qevols}
q_{j}^{'}= wq_{j} + S \,,
\end{equation}
with $S$ being the sum of the terms on the right hand side 
of (\ref{qevol}) excluding the $j$-th term.

For $t$ between the grid points $t_{j-1}$ and $t_{j}$ we interpolate 
the parton density using quadratic interpolation as follows
%(A.22)
\begin{equation}
q(x_{i},t)=at^{2} + bt + c \,.
\end{equation}
Thus we relate the value of $q$ at the point $t_j$ to that of $q$
at the point $t_{j-1}$ by
%(A.23)
\begin{equation}
\label{qsolv}
q(x_{i},t_{j})=q(x_{i},t_{j-1})+\frac{1}{2}
[q^{'}(x_{i},t_{j})+q^{'}(x_{i},t_{j-1})]\Delta t_{j} \,,
\end{equation}
where \( \Delta t_{j}=t_{j}-t_{j-1} \). This equation can also be written 
more compactly as
\begin{equation}
\label{ipol}
q_j=q_{j-1}+\frac{1}{2}(q^{'}_{j-1}+q^{'}_{j})\Delta t_j \,.
\end{equation}
The resulting system of two linear equations (\ref{ipol}) 
and (\ref{qevols}) for $q_j$ and $q^{'}_j$ has the solution 
%(A.24)
\begin{equation}
q_j=\frac{2q_{j-1}+(q^{'}_{j-1}+S)\Delta t_j}{2-w\Delta t_j} \,,
\end{equation}
and yields $q^{'}_j$ from (\ref{qevols}). 
Applying the same procedure to the gluon and singlet equations 
Eqs.(A.10)-(A.11) involves four equations because we have to compute both the 
densities and their derivatives.

The evolution proceeds from the initial $\mu_0^2 = \mu^2_{\rm LO}$ (or 
$\mu_0^2 = \mu^2_{\rm NLO}$) 
to the first matching point at the scale $\mu^2 = m_c^2$. Next 
the charm density is introduced in the 
NNLO ($\alpha^{2}_{s}$-order terms) and all the four-flavor
densities are evolved from the boundary conditions in 
Ens.(\ref{eqn2.8})-(\ref{eqn2.10}). This evolution continues up to the  
transition point $\mu^2 = m_b^2$, where the same procedure 
is applied to generate the bottom quark density.
At that matching point all five-flavor densities are evolved 
starting from the boundary conditions in Eqs.(\ref{eqn2.8})-(\ref{eqn2.11})
up to all higher $\mu^2$ scales.
%\end{document}

%\documentstyle[12pt]{article}
%\pagestyle{myheadings}  
%\begin{document}
%----------------------------References-------------------------------------
%

%\documentstyle[12pt]{article}
%\begin{document}
\centerline{\bf \large{Figure Captions}}
\begin{description}
%--------------------------------------
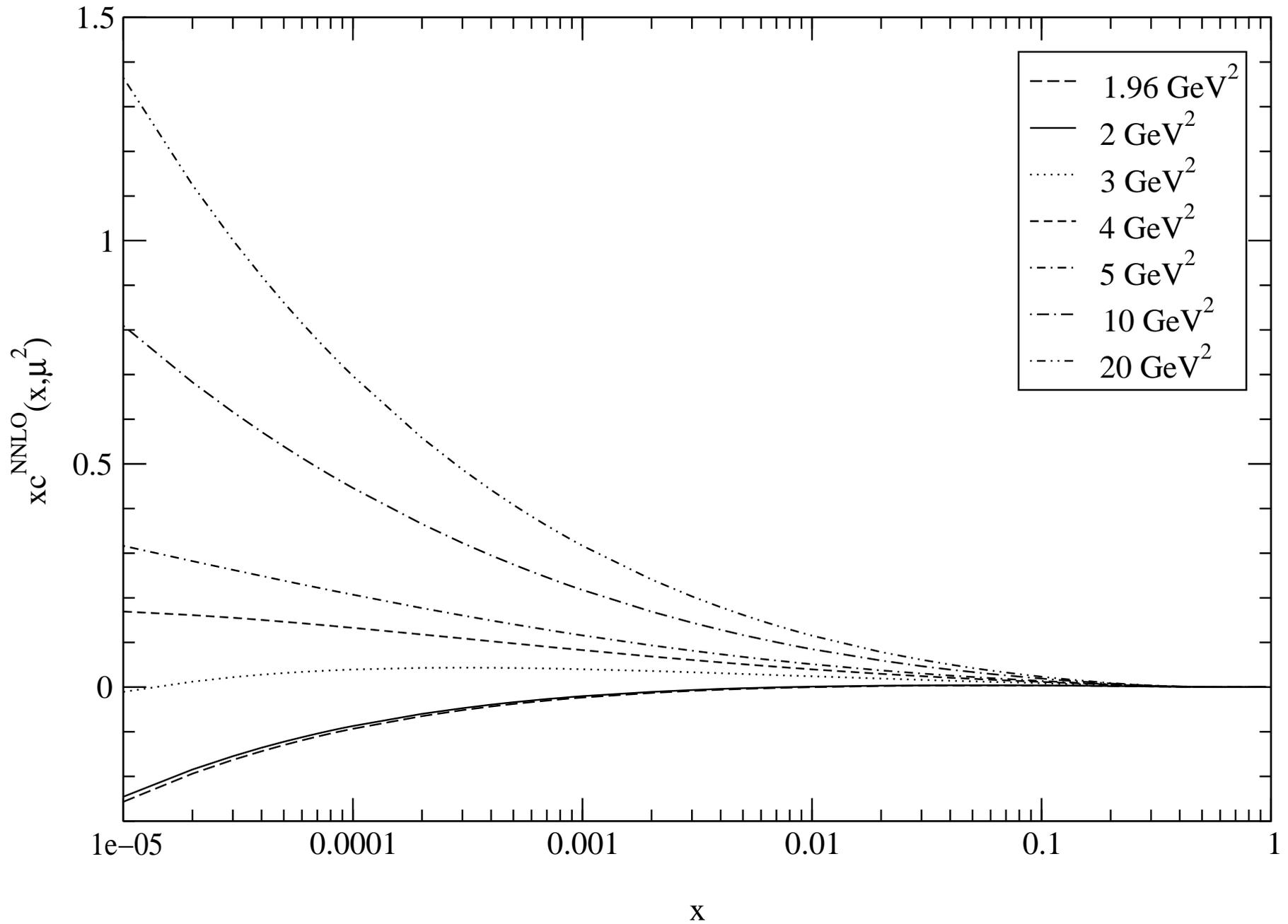
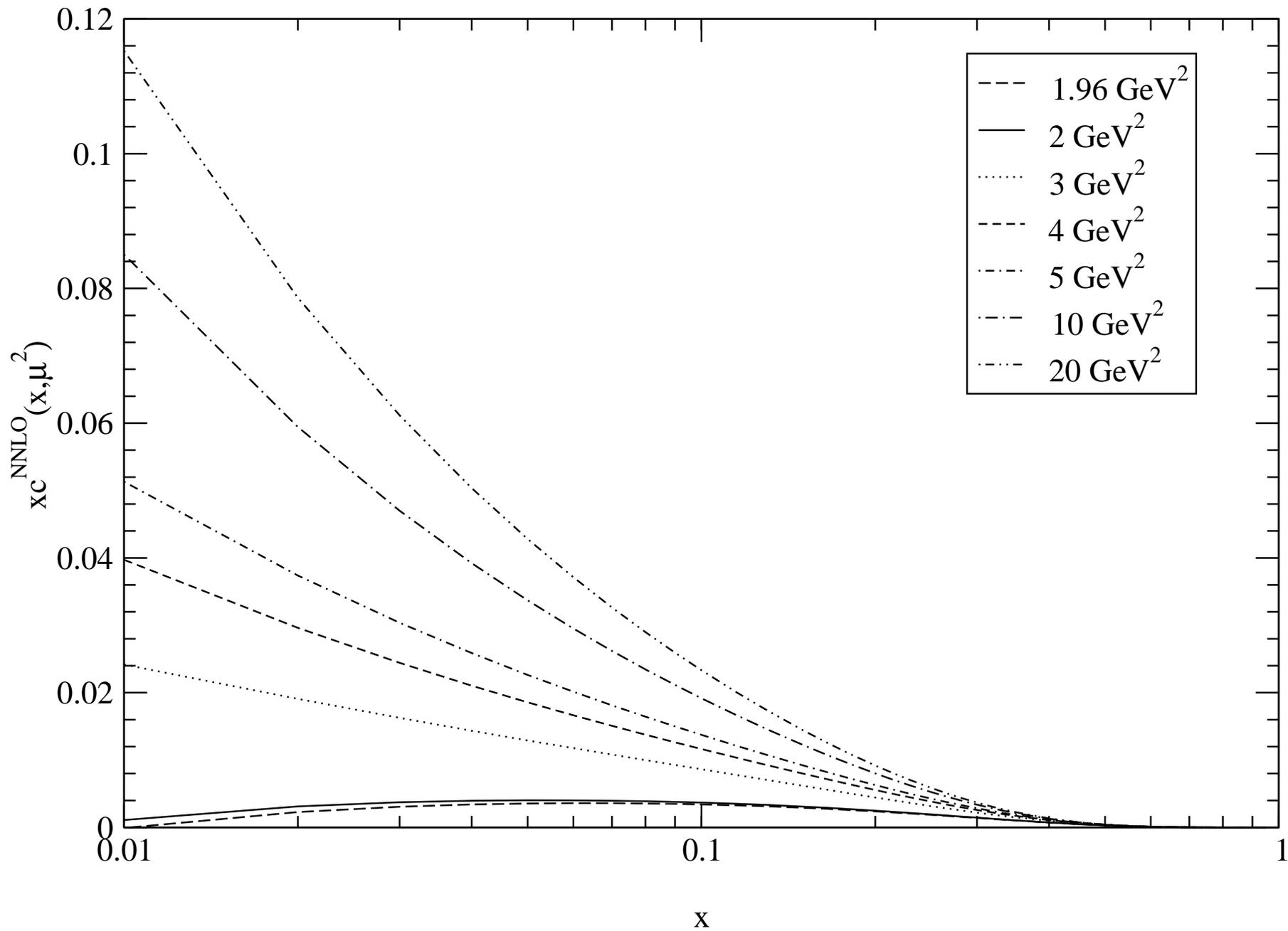
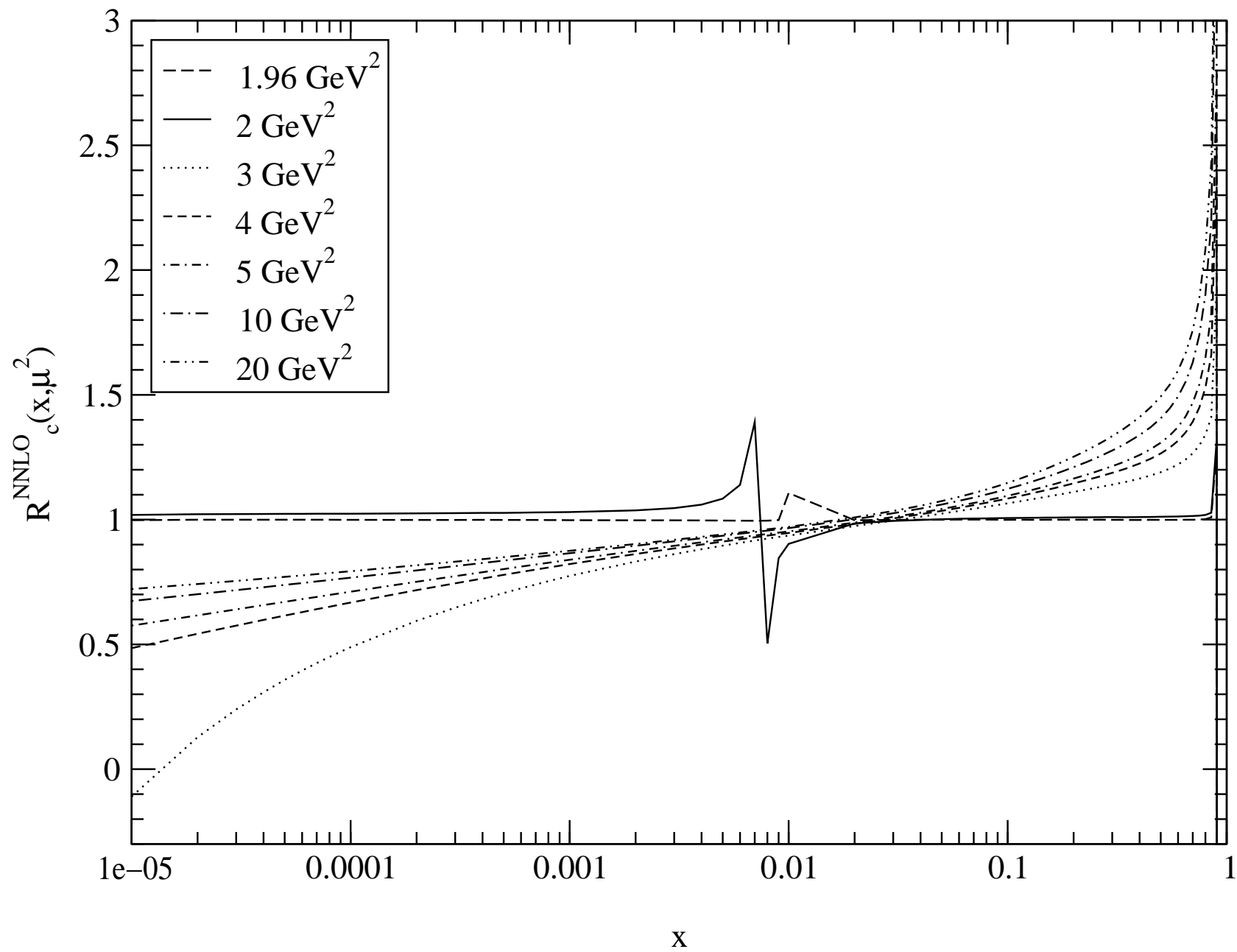
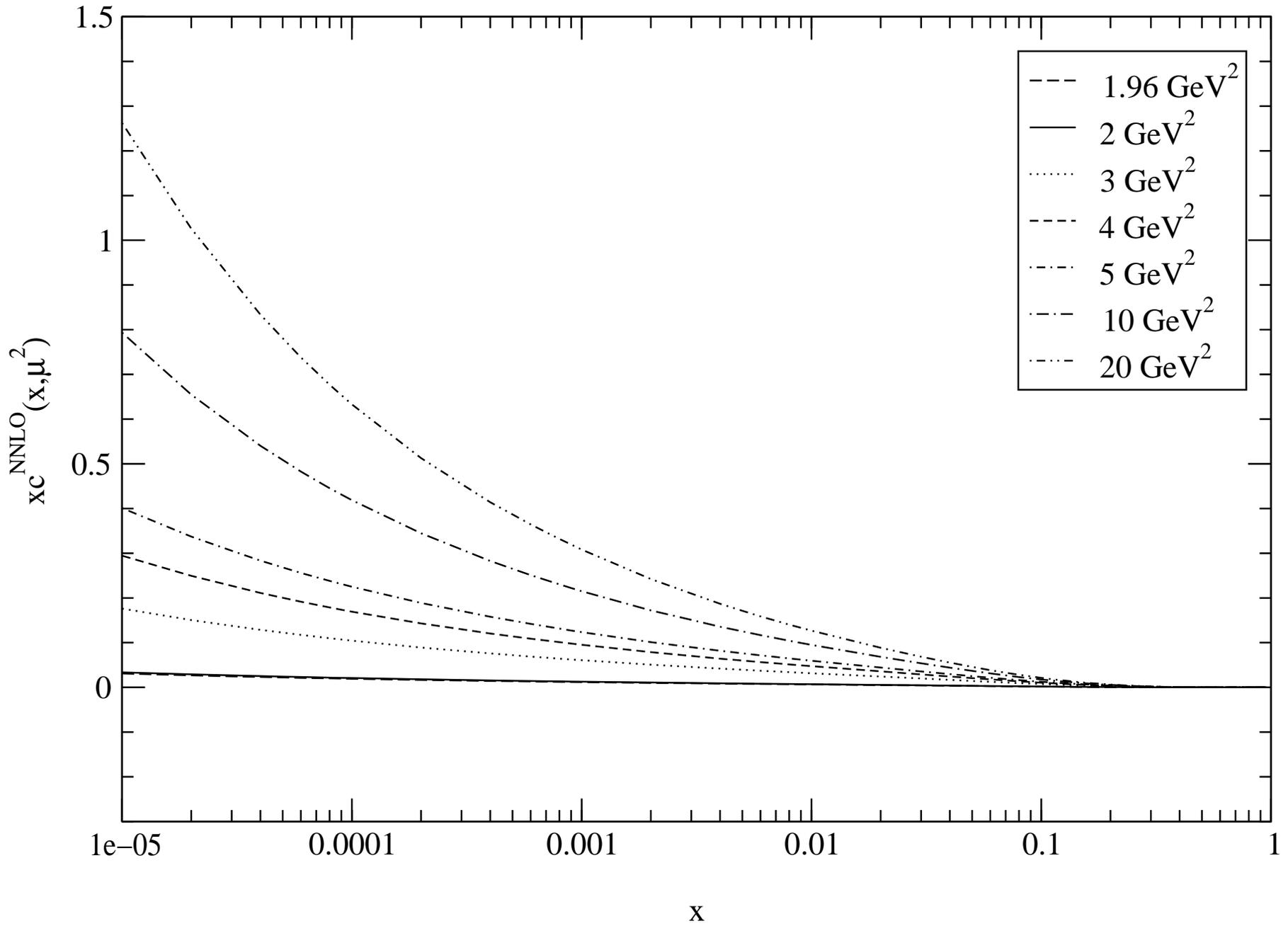
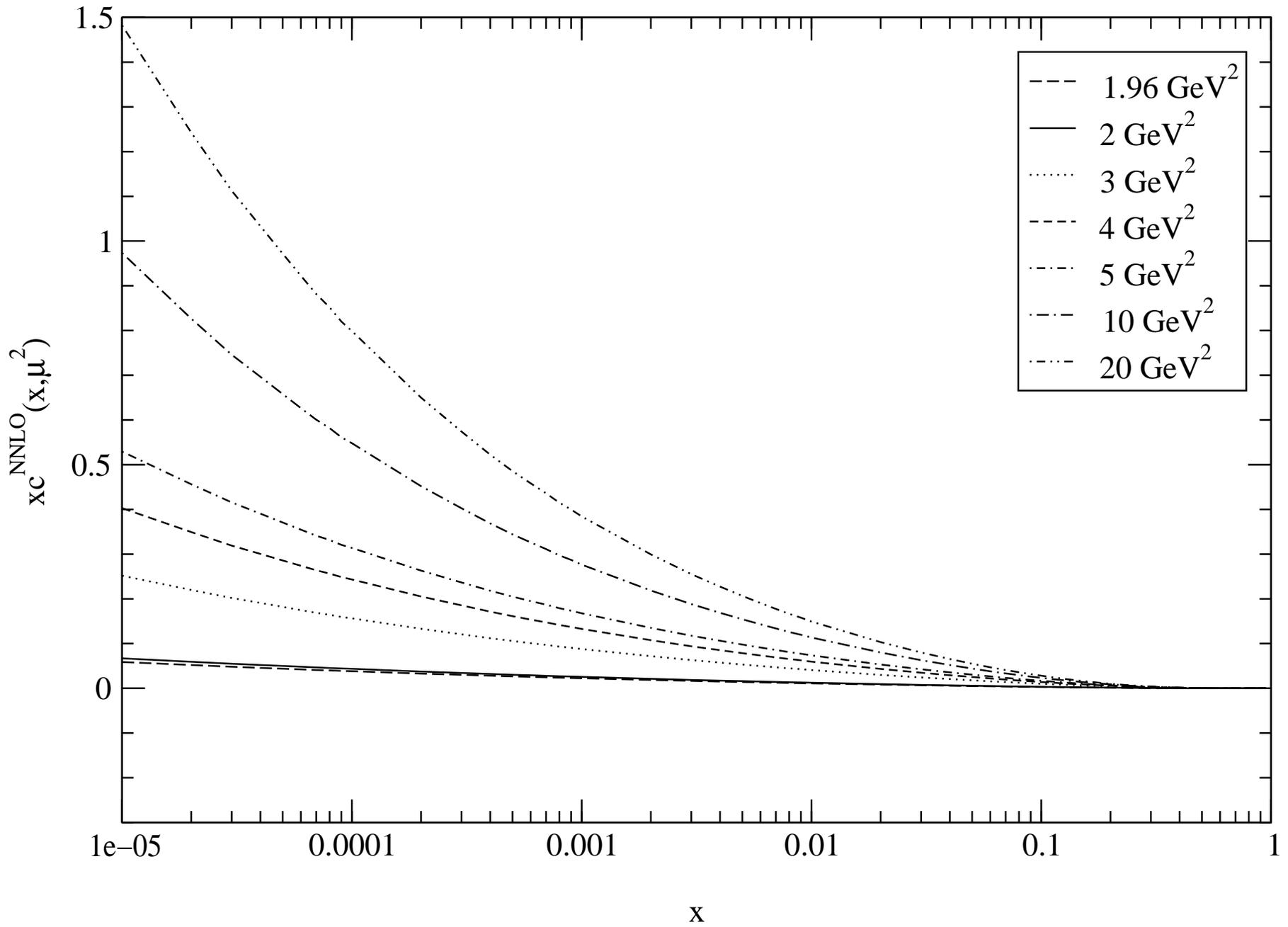
\item[Fig. 1.]
(a) The charm quark density $xc_{\rm NNLO}(4,x,\mu^2)$ the range
$10^{-5} < x < 1$ for 
$\mu^2 =$ 1.96, 2, 3, 4, 5, 10 and 20 in units 
of $({\rm GeV/c}^2)^2$, 
(b) similar plot as in (a) but now for $0.01 < x < 1$,
(c) ratios 
$R^{\rm NNLO}_{c}(x,\mu^2)=xc_{\rm EVOLVED}(4,x,\mu^2)$/$xc_{\rm FOPT}(4,x,\mu^2)$ 
for the same scales, 
(d) and (e) the NLO results from MRST98 set 1 and CTEQ5HQ respectively.
%----------------------
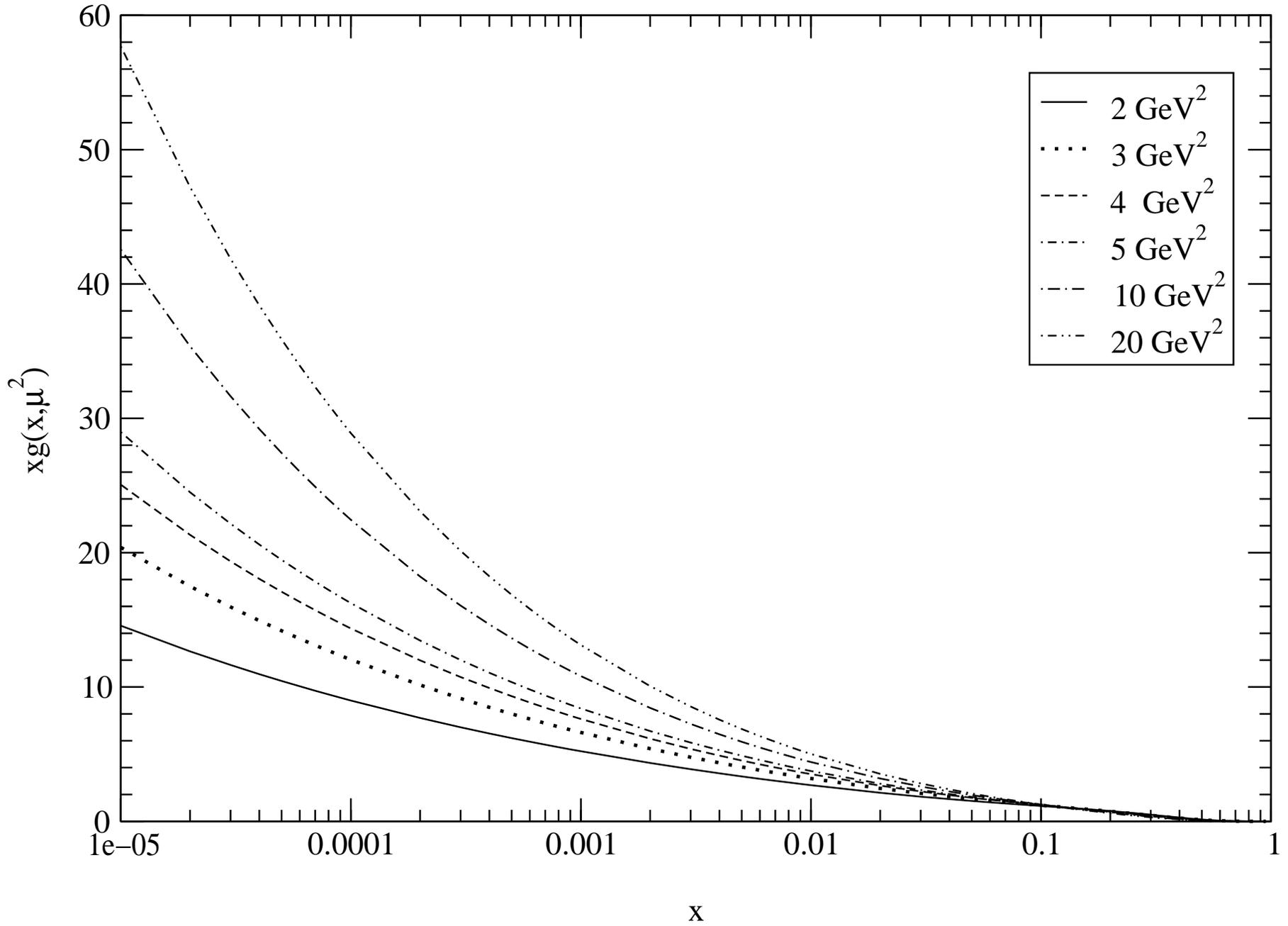
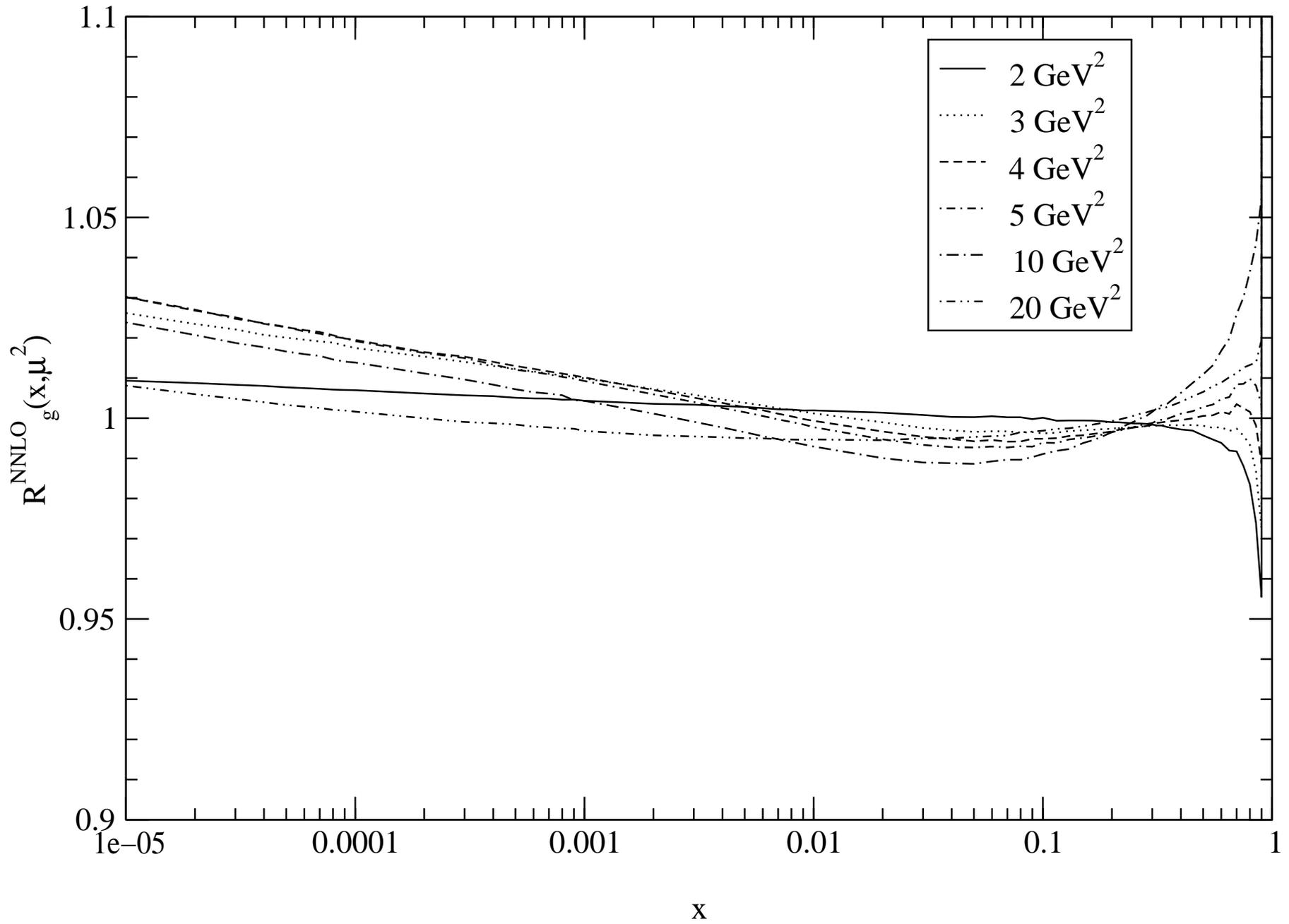
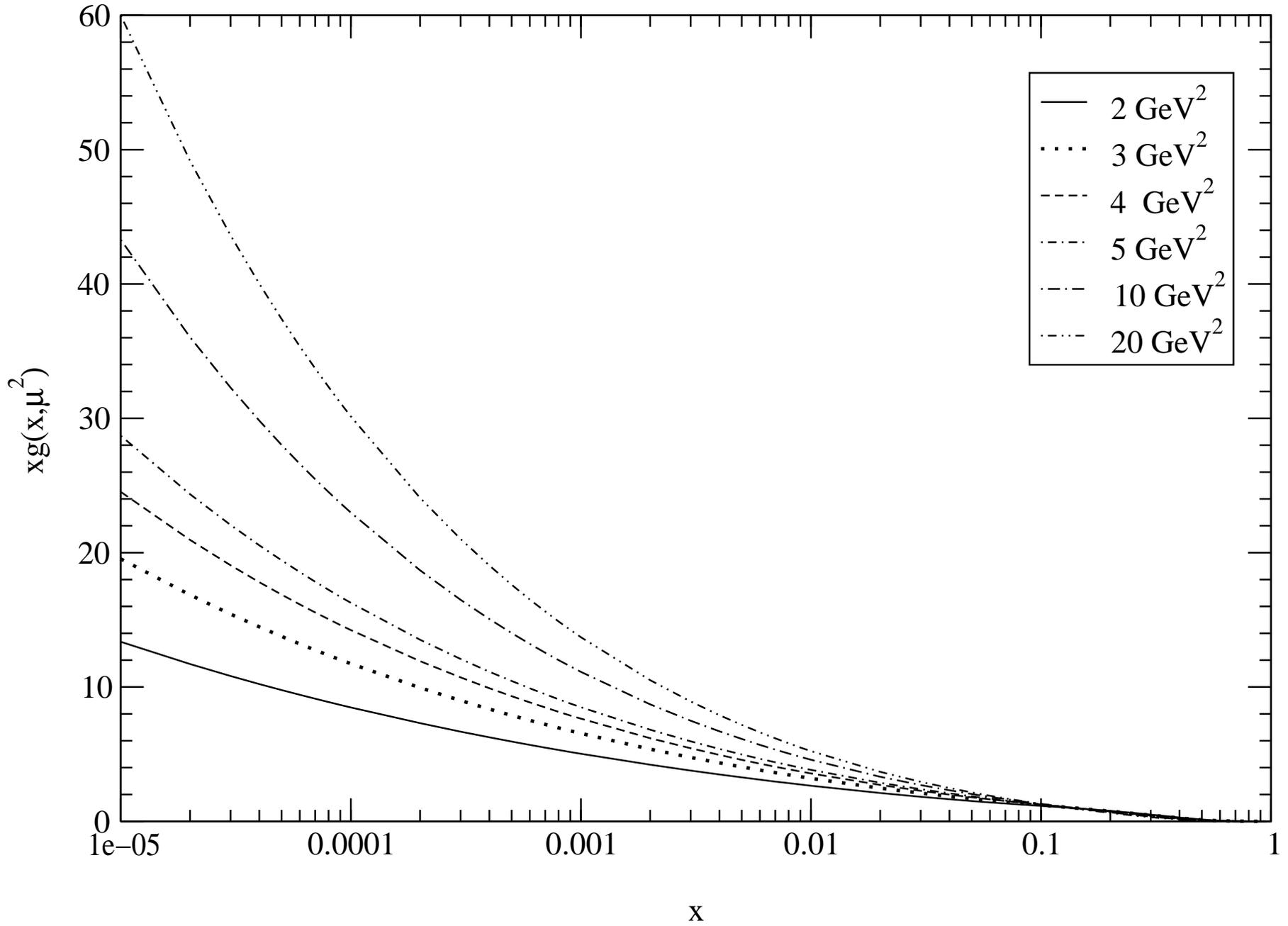
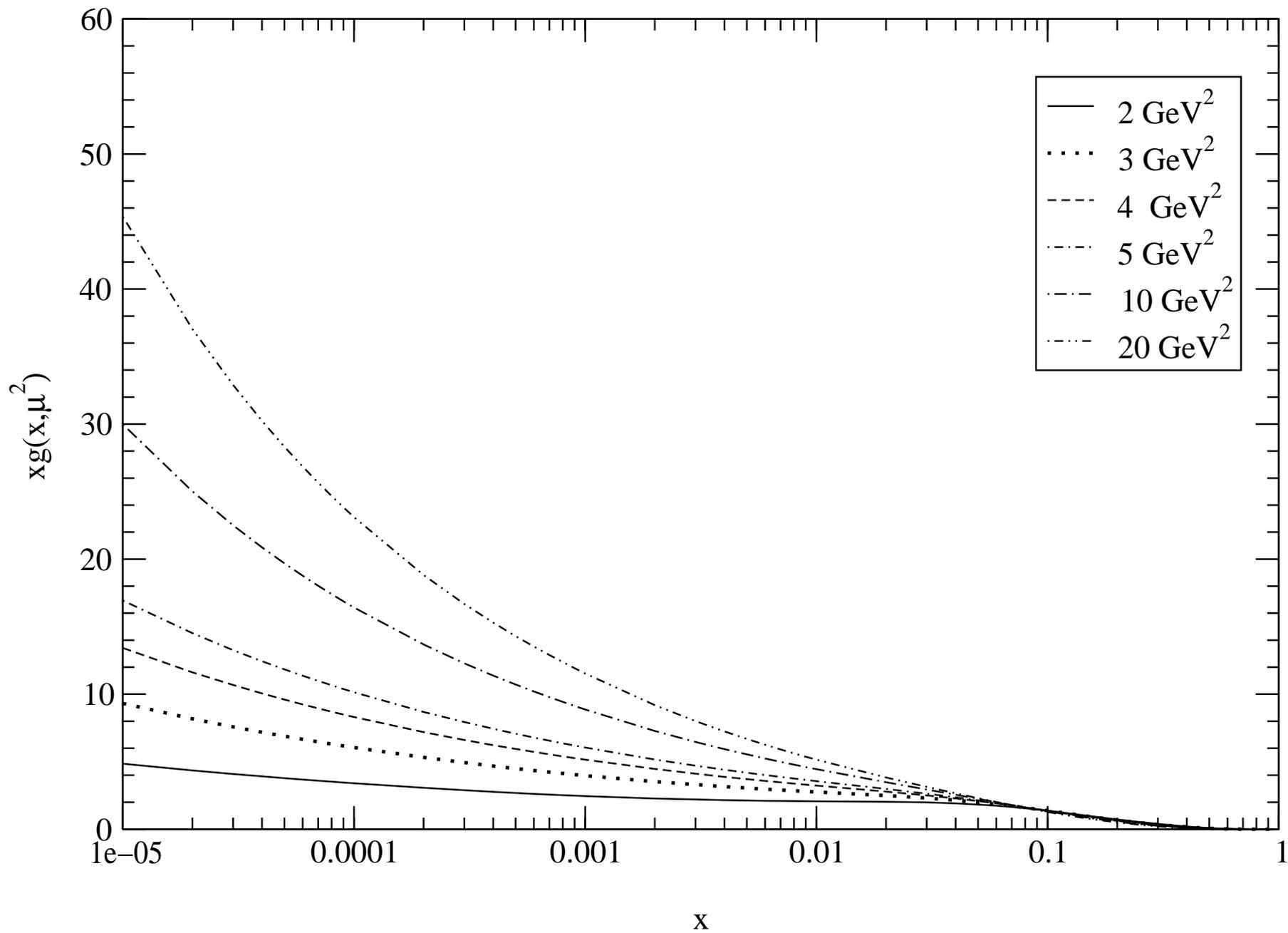
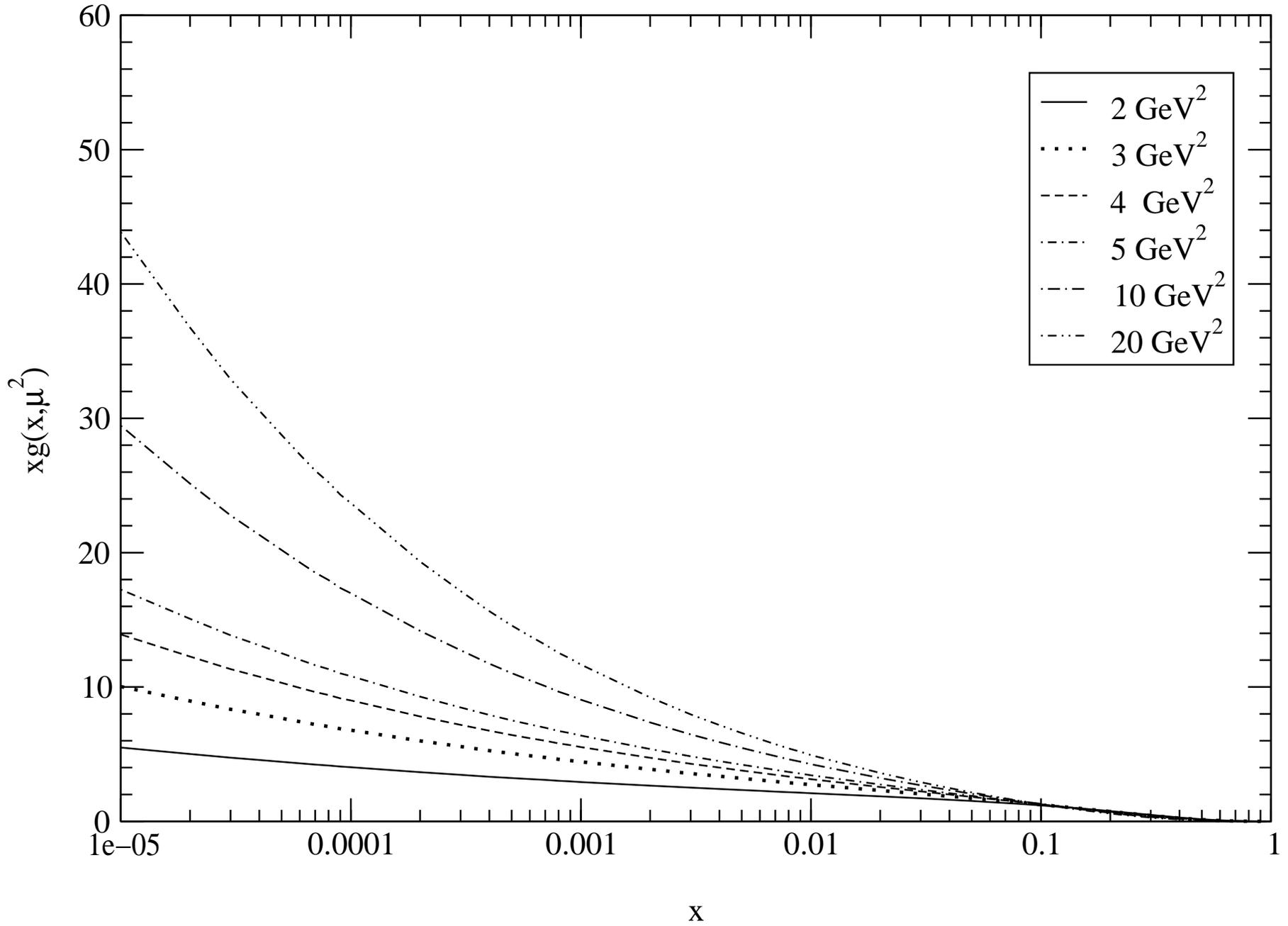
\item[Fig. 2.]
(a) The gluon density $xg_{\rm NNLO}(4,x,\mu^2)$ in the range
$10^{-5} < x < 1$ for 
$\mu^2 =$ 2, 3, 4, 5, 10 and 20 in units 
of $({\rm GeV/c}^2)^2$, 
(b) ratios 
$R^{\rm NNLO}_{g}(x,\mu^2)=xg_{\rm EVOLVED}(4,x,\mu^2)$/$xg_{\rm FOPT}(4,x,\mu^2)$ 
for the same scales,  
(c) the three-flavor NLO gluon density in the same range,
(d) and (e) the NLO results from MRST98 set 1 and CTEQ5HQ respectively.
%--------------------------------------------
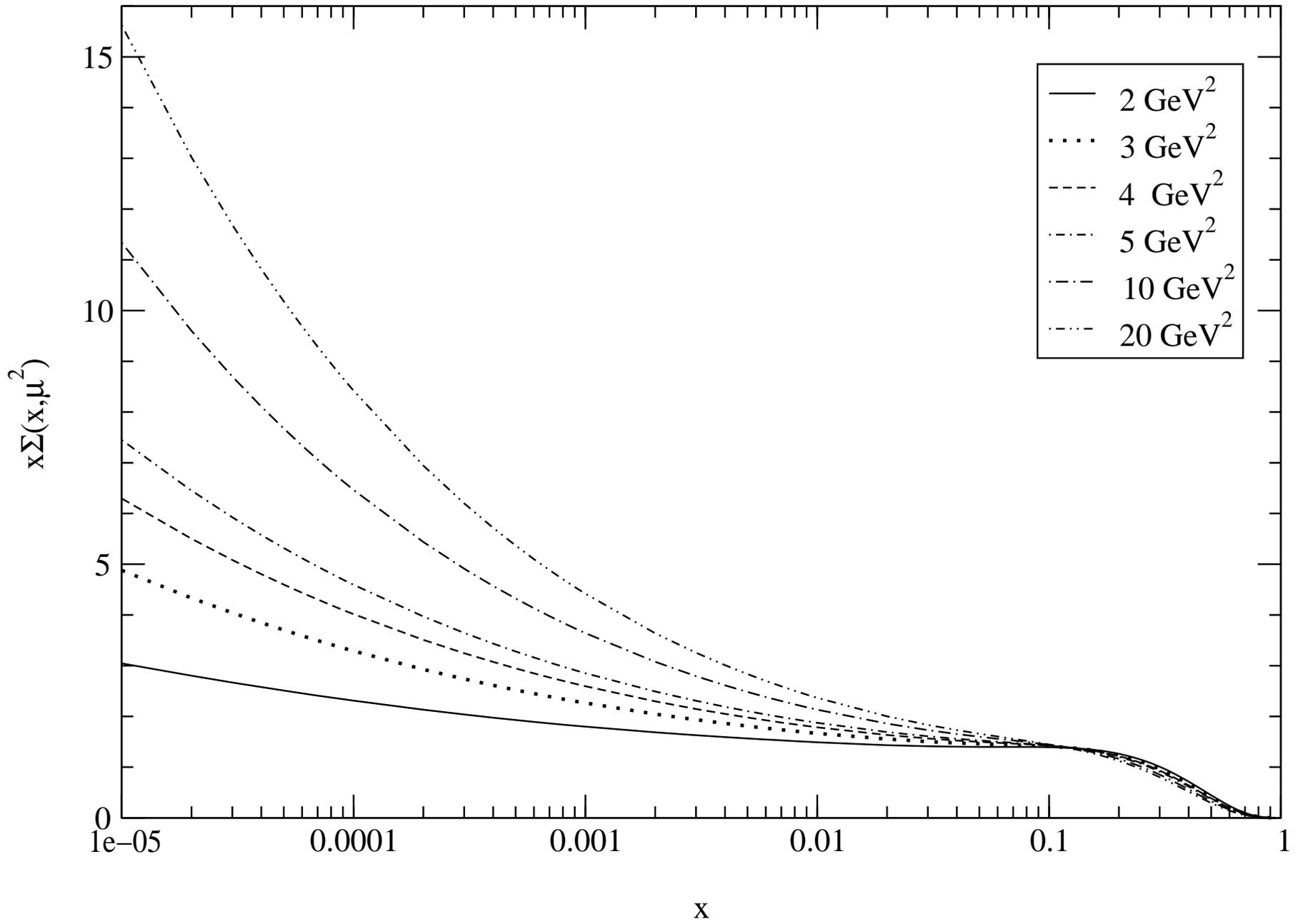
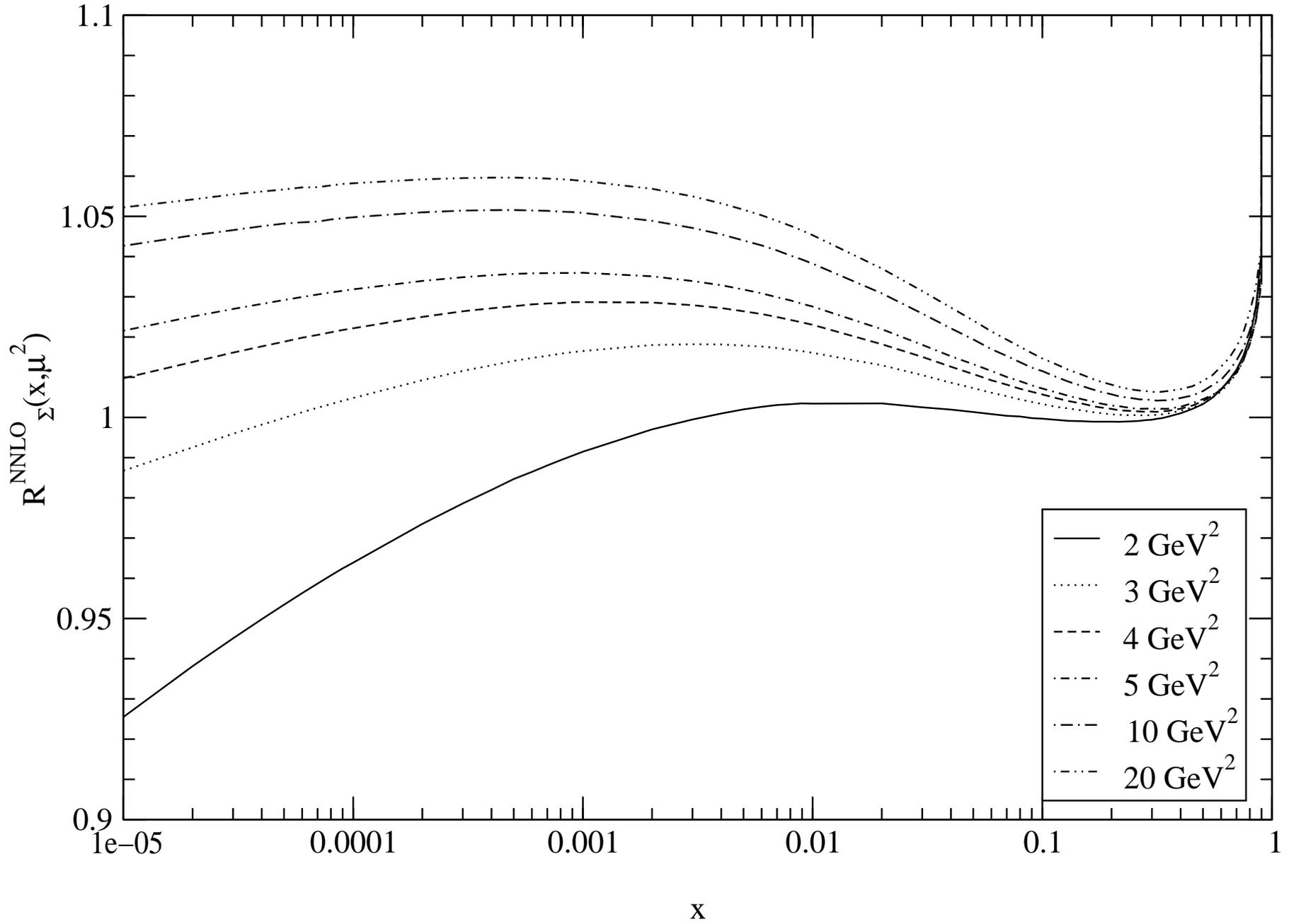
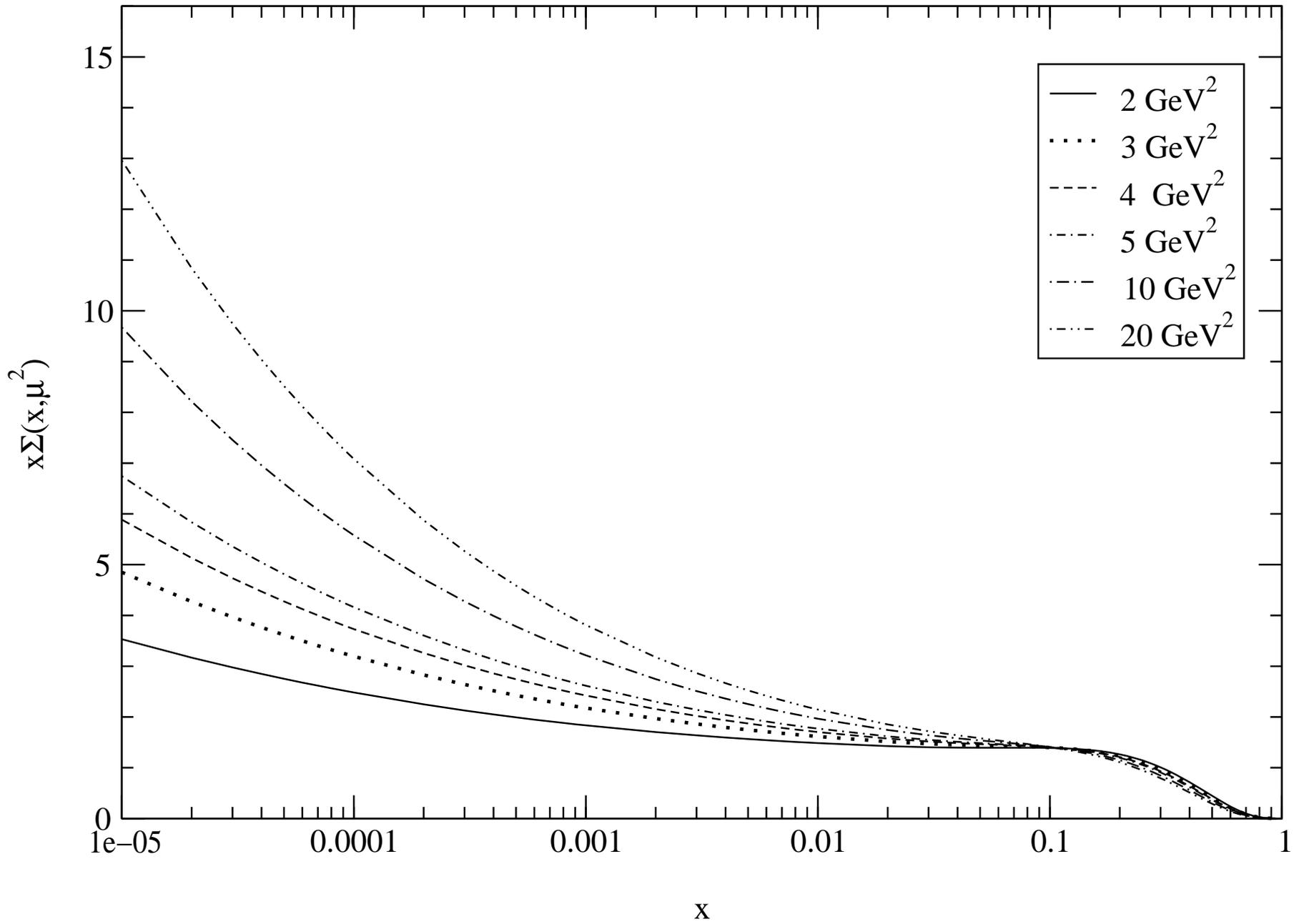
\item[Fig. 3.]
(a) The singlet density $x\Sigma_{\rm NNLO}(4,x,\mu^2)$ 
in the range $10^{-5} < x < 1$ for 
$\mu^2 =$ 2, 3, 4, 5, 10 and 20 in units 
of $({\rm GeV/c}^2)^2$, 
(b) ratios 
$R^{\rm NNLO}_{\Sigma}(x,\mu^2)=x\Sigma_{\rm EVOLVED}(4,x,\mu^2)$
/$x\Sigma_{\rm FOPT}(4,x,\mu^2)$ for the same scales,
(c) the three-flavor NLO density.
%----------------------------------------
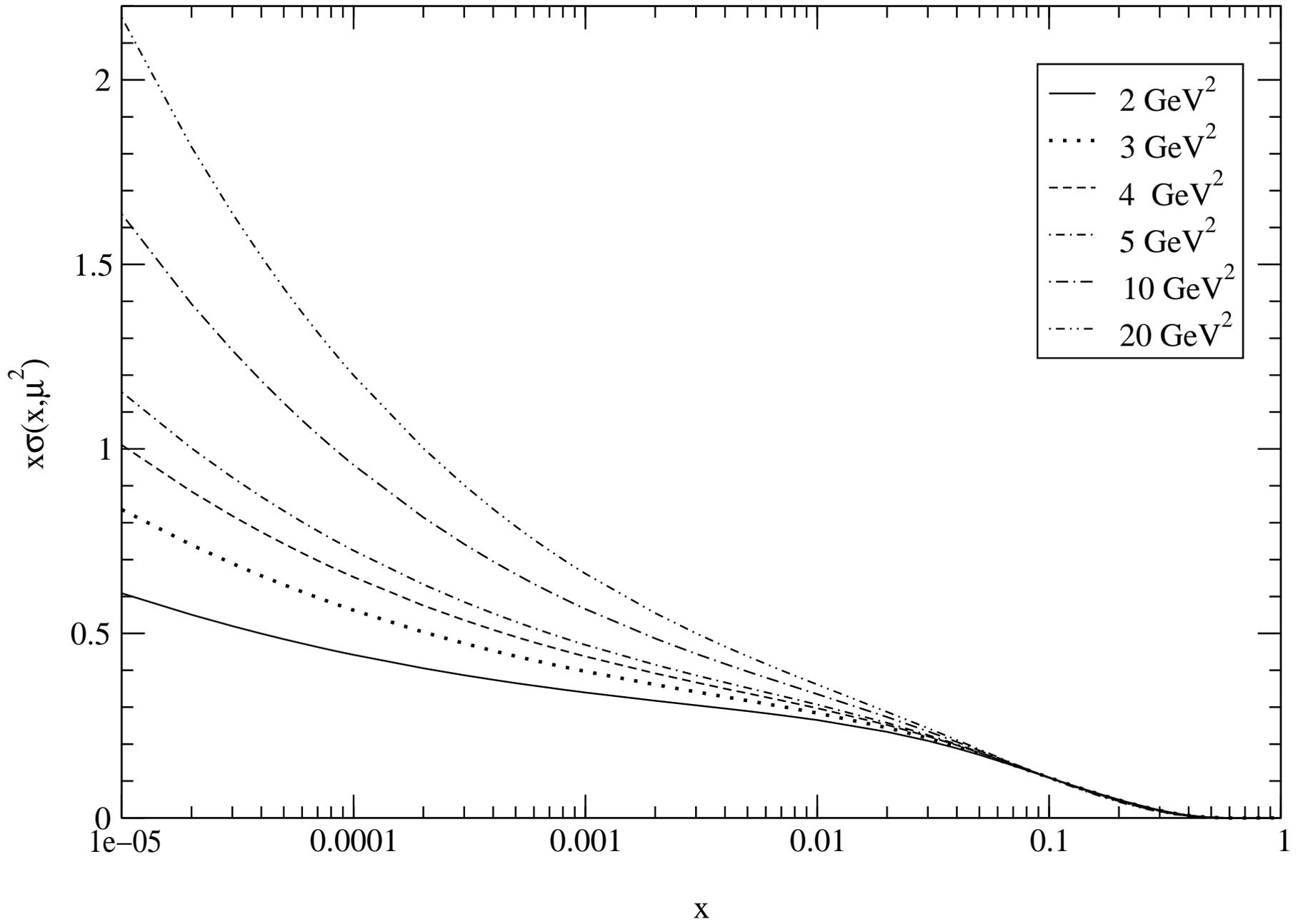
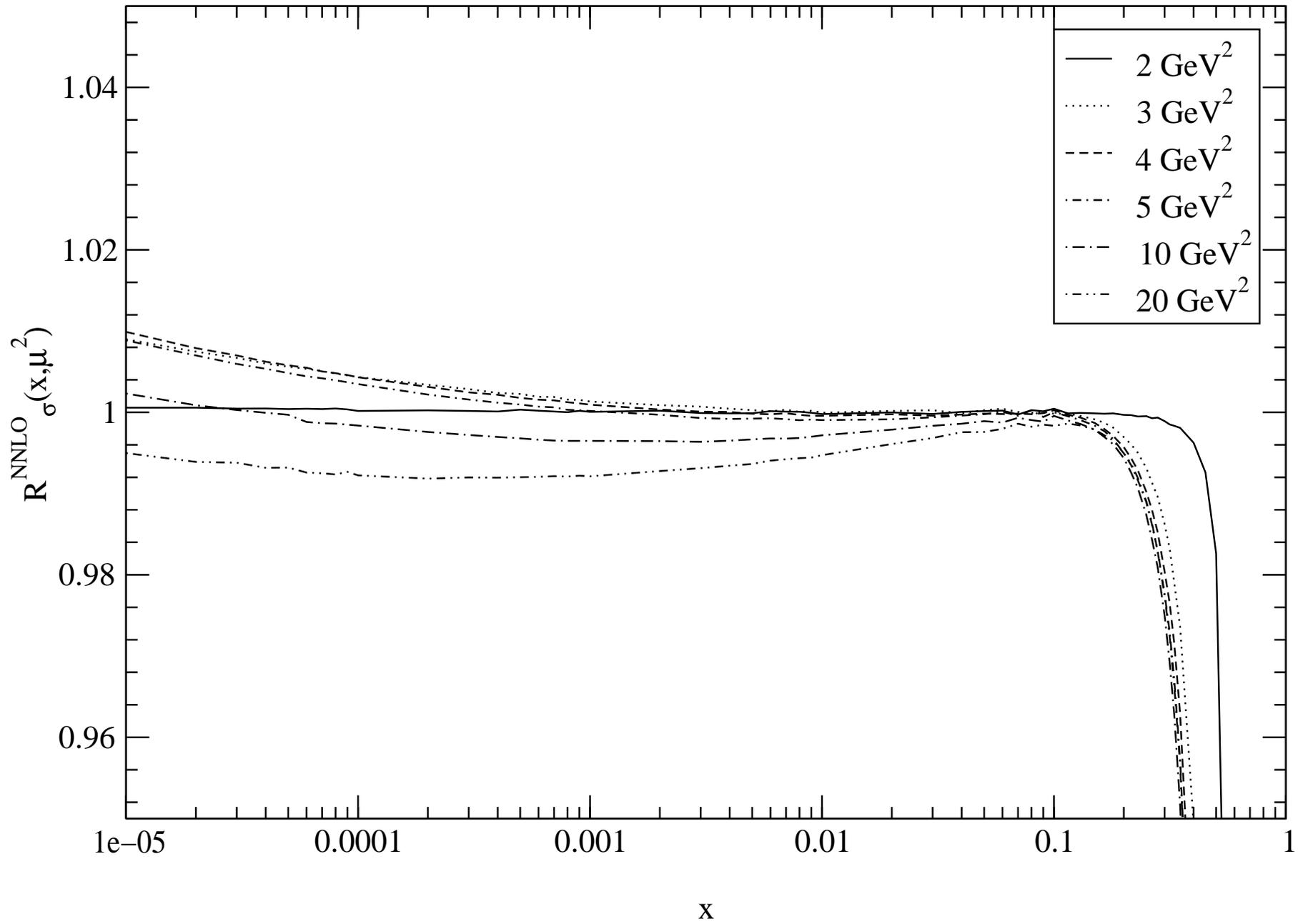
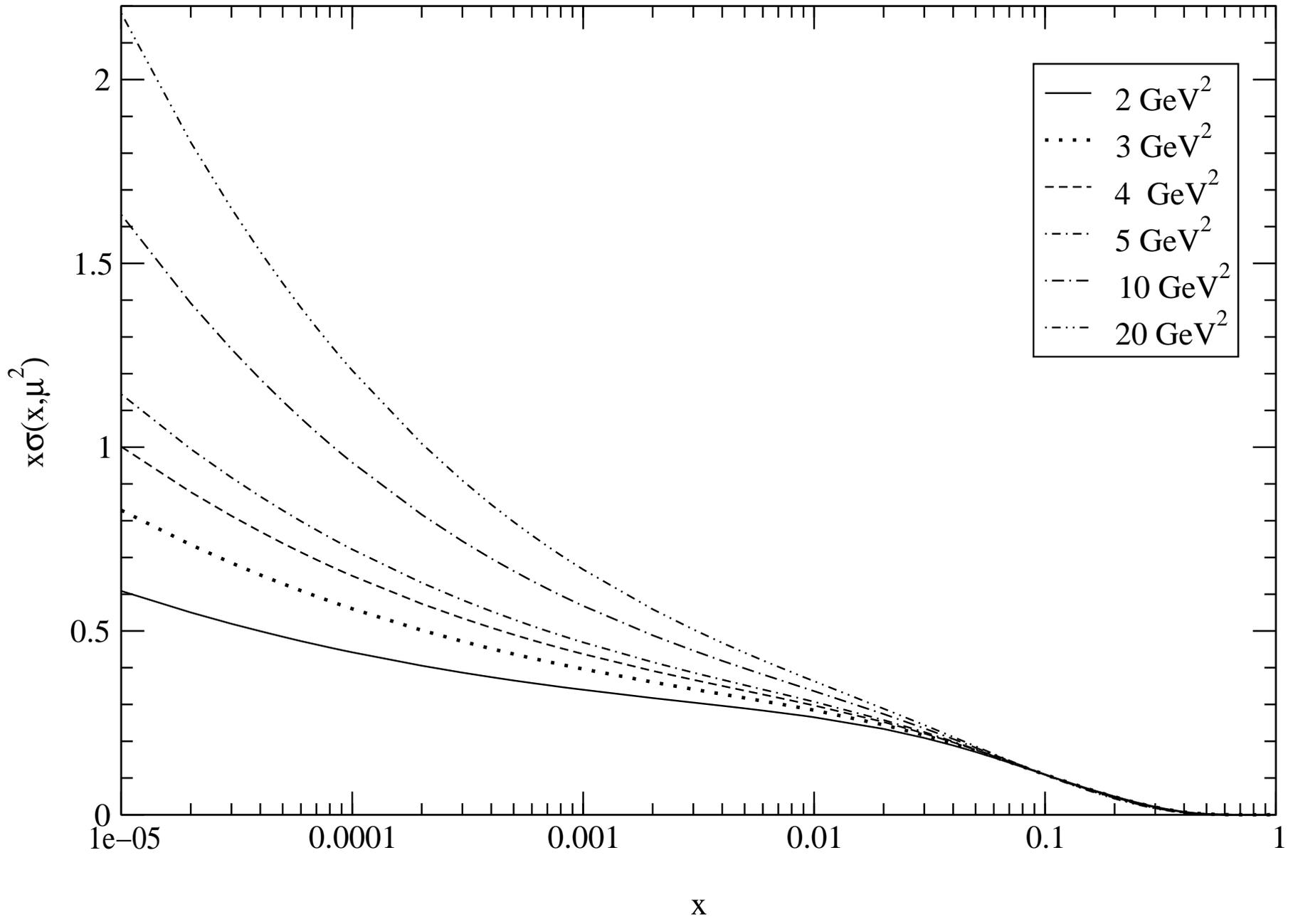
\item[Fig. 4.]
(a) The nonsinglet quark density $x\sigma_{\rm NNLO}(4,x,\mu^2)$  
in the range $10^{-5} < x < 1$ for 
$\mu^2 =$ 2, 3, 4, 5, 10 and 20 in units 
of $({\rm GeV/c}^2)^2$, 
(b) ratios 
$R^{\rm NNLO}_{\sigma}(x,\mu^2)=x\sigma_{\rm EVOLVED}(4,x,\mu^2)$/$x\sigma_{\rm FOPT}(4,x,\mu^2)$ 
for the same scales,
(c) the three-flavor NLO density.
%---------------------------------
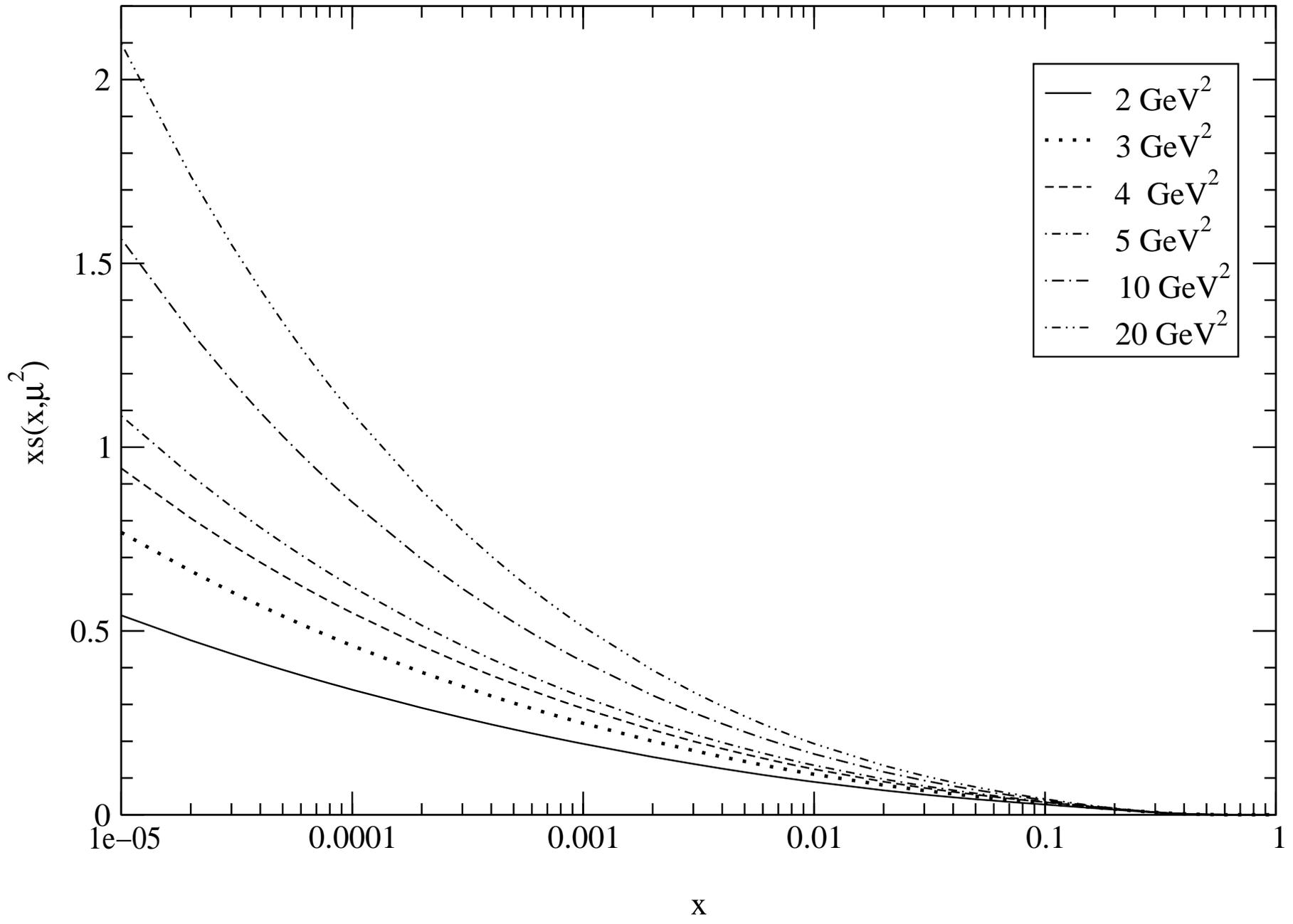
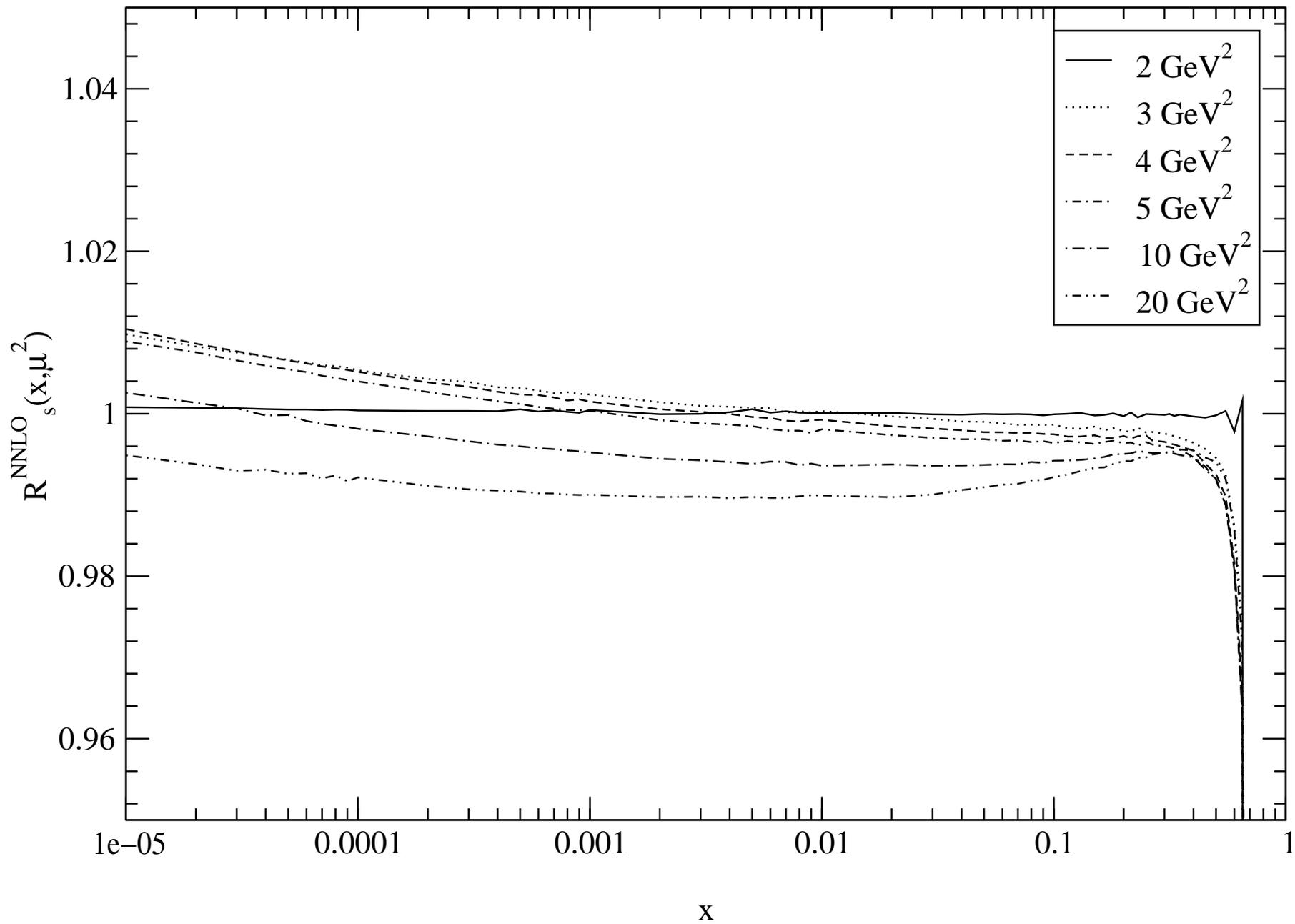
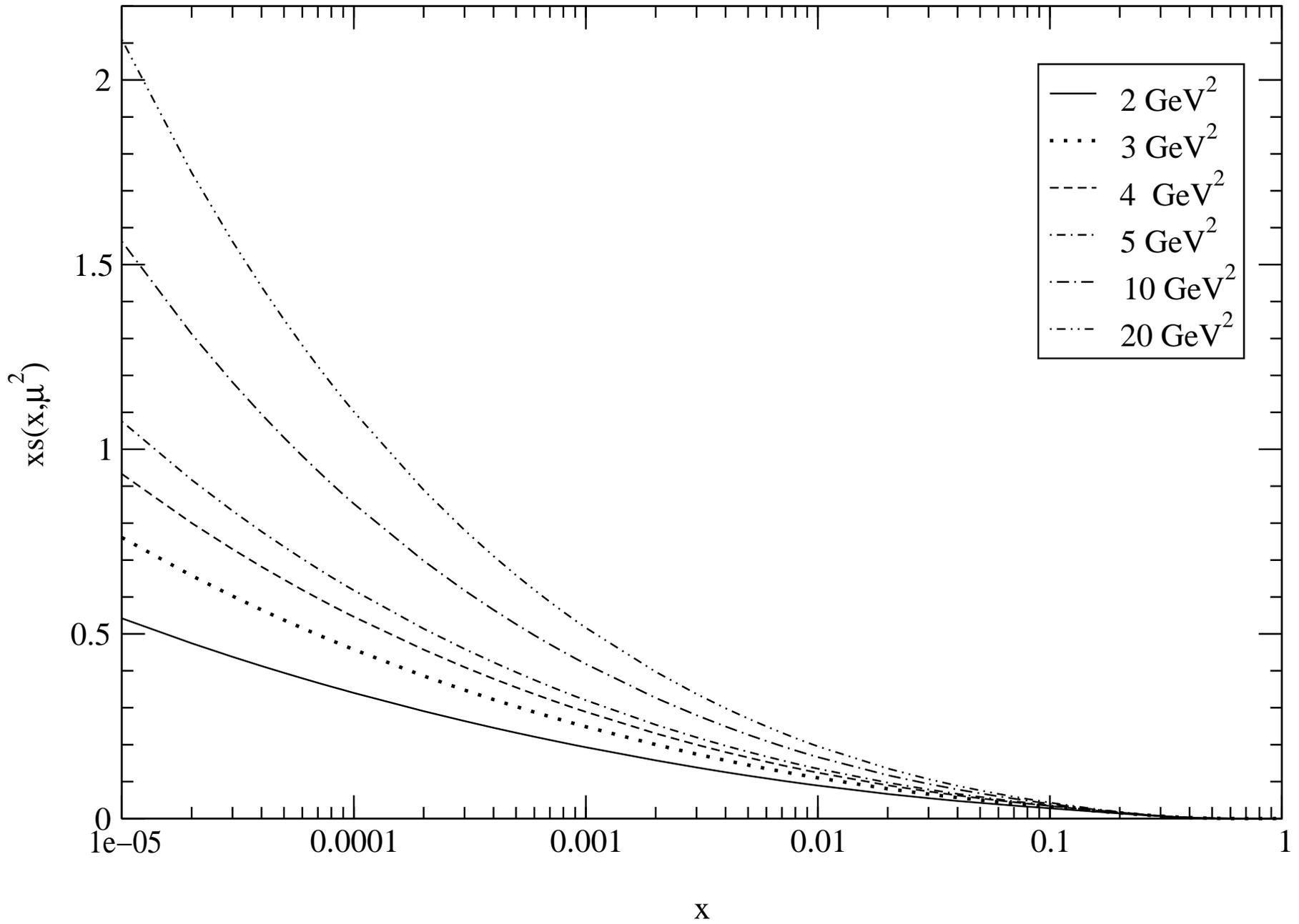
\item[Fig. 5.]
(a) The strange quark density $xs_{\rm NNLO}(4,x,\mu^2)$ in the range
$10^{-5} < x < 1$ for 
$\mu^2 =$ 2, 3, 4, 5, 10 and 20 in units 
of $({\rm GeV/c}^2)^2$, 
(b) ratios 
$R^{\rm NNLO}_{s}(x,\mu^2)=xc_{\rm EVOLVED}(4,x,\mu^2)$/$xc_{\rm FOPT}(4,x,\mu^2)$ 
for the same scales,
(c) the three-flavor NLO density.
%-------------------------------
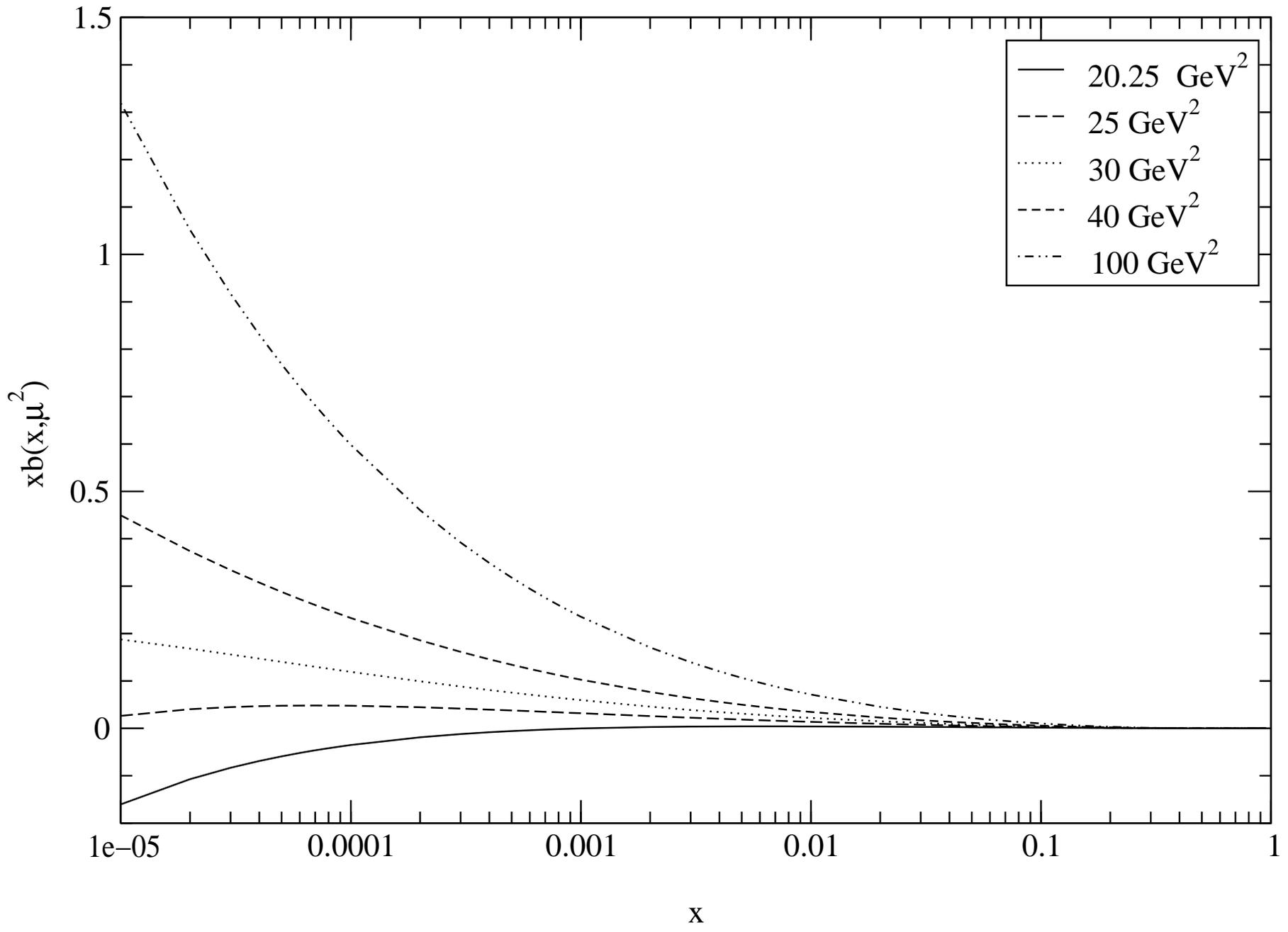
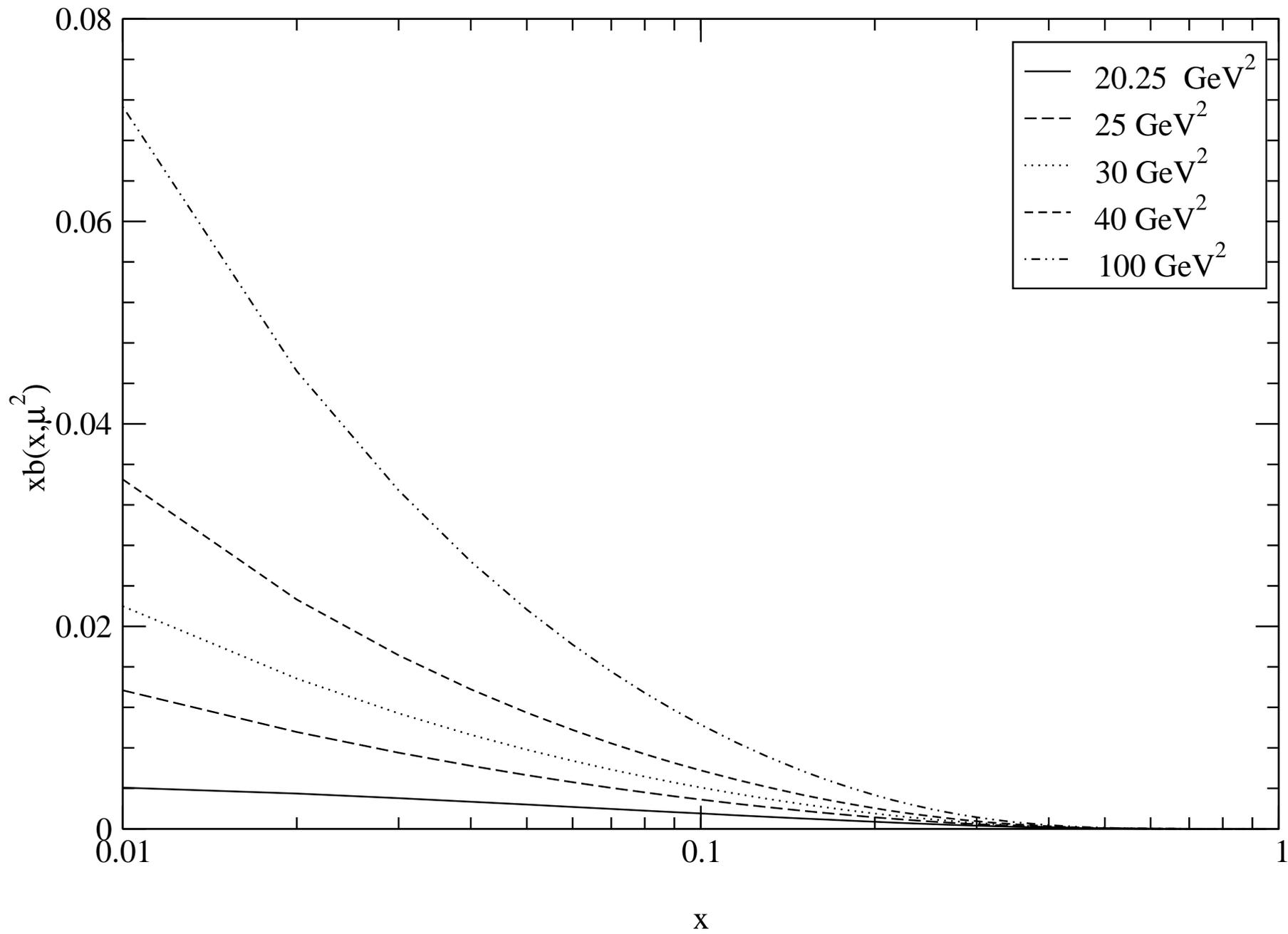
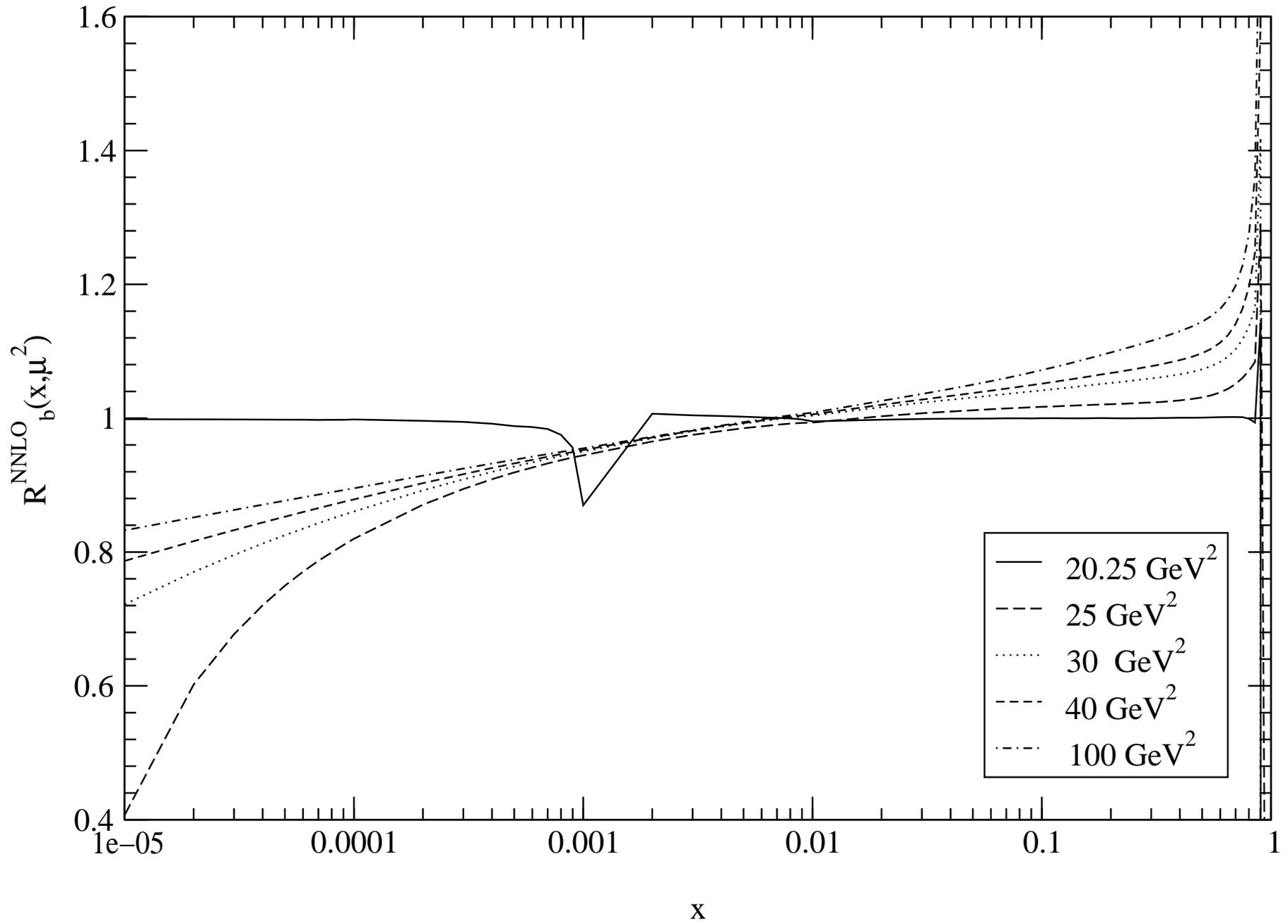
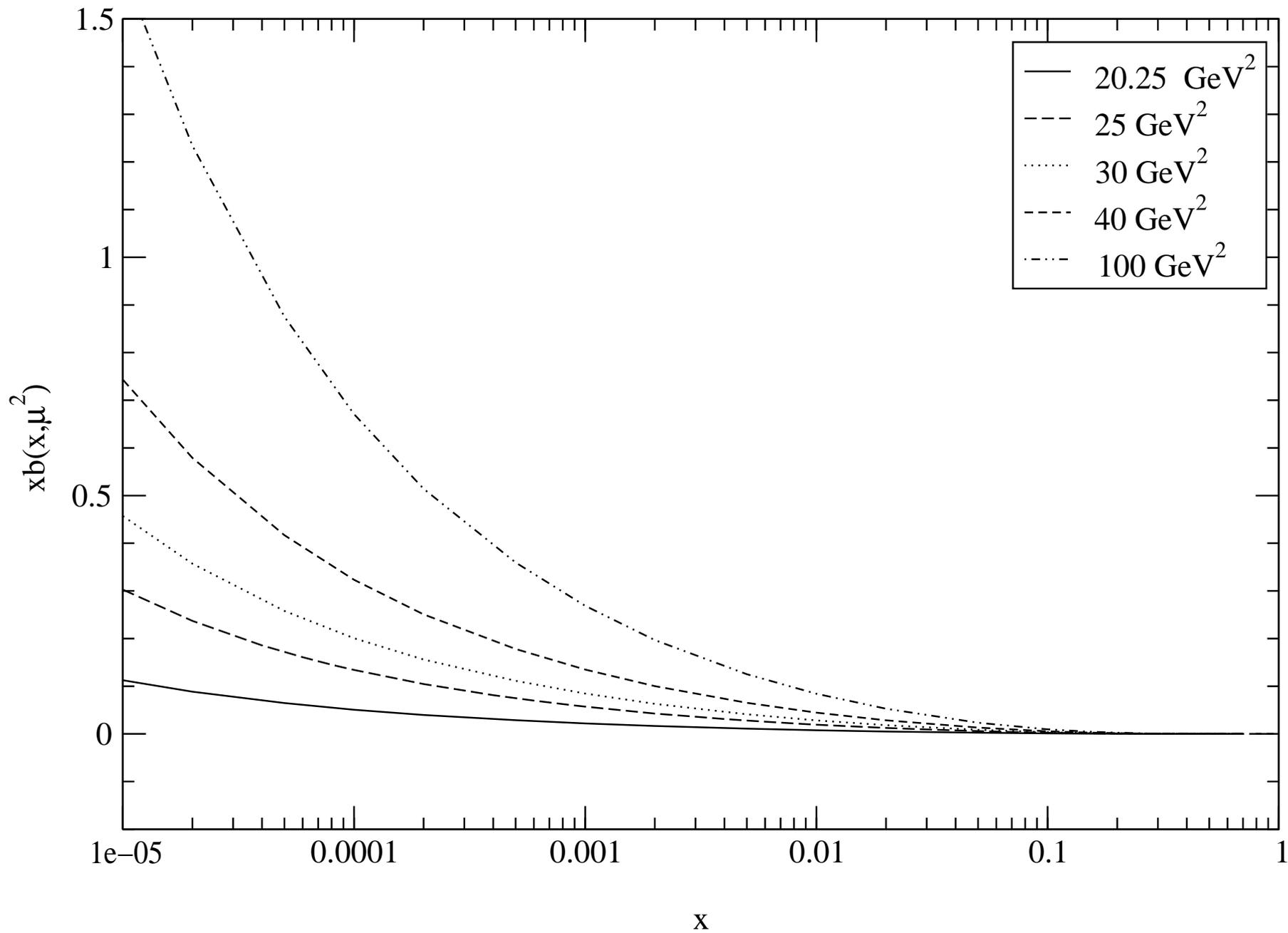
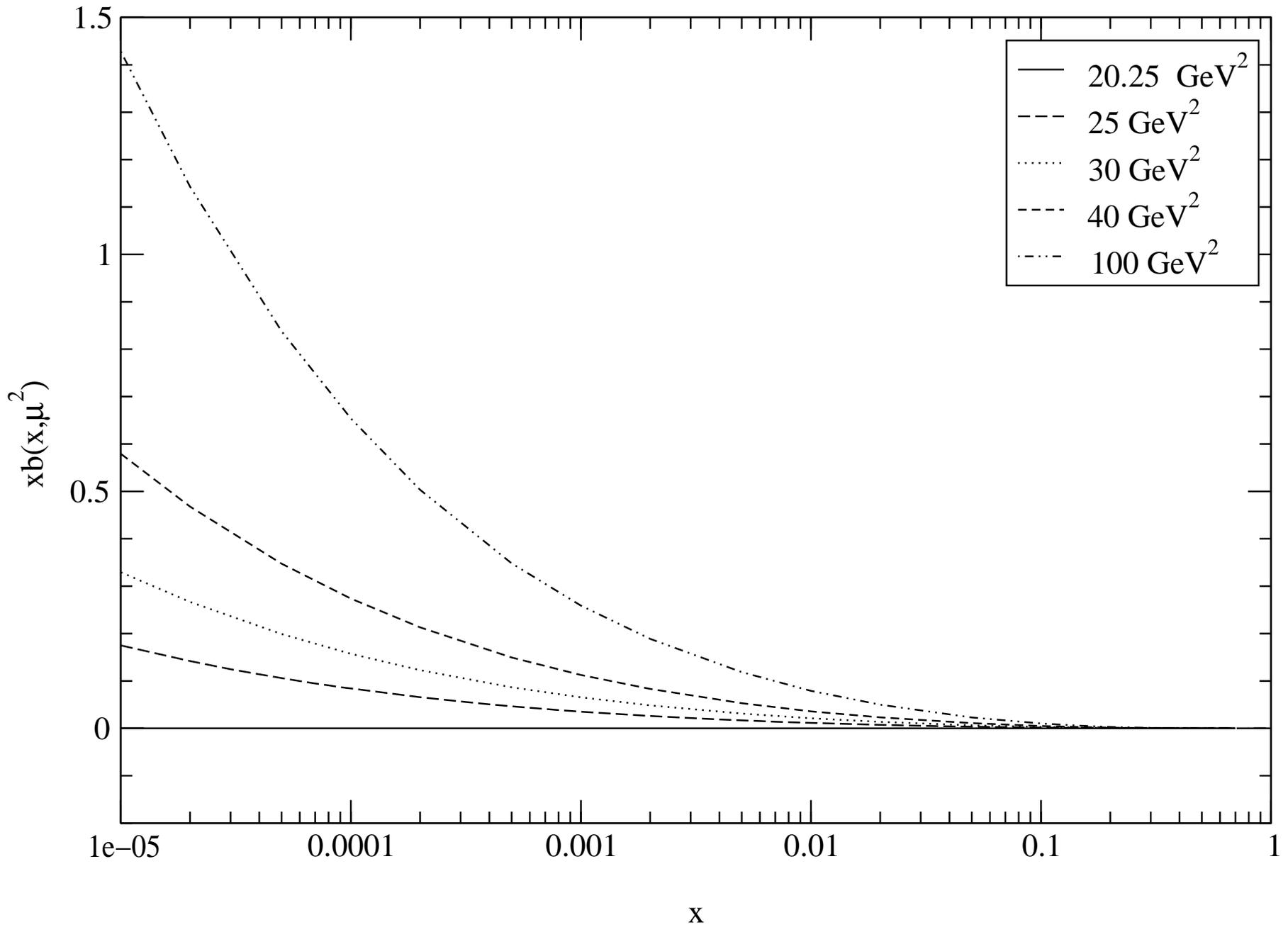
\item[Fig. 6.]
(a) The bottom quark density $xb_{\rm NNLO}(5,x,\mu^2)$ in the range
$10^{-5} < x < 1$ for 
$\mu^2 =$ 20.25, 25, 30, 40 and 100 in units 
of $({\rm GeV/c}^2)^2$, 
(b) similar plot as in (a) but now for $0.01 < x < 1$,
(c) ratios 
$R^{\rm NNLO}_{b}(x,\mu^2)=xb_{\rm EVOLVED}(4,x,\mu^2)$/$xb_{\rm FOPT}(4,x,\mu^2)$ 
for the same scales,
(d) and (e) the NLO results from MRST98 set 1 and CTEQ5HQ respectively.
%----------------------------
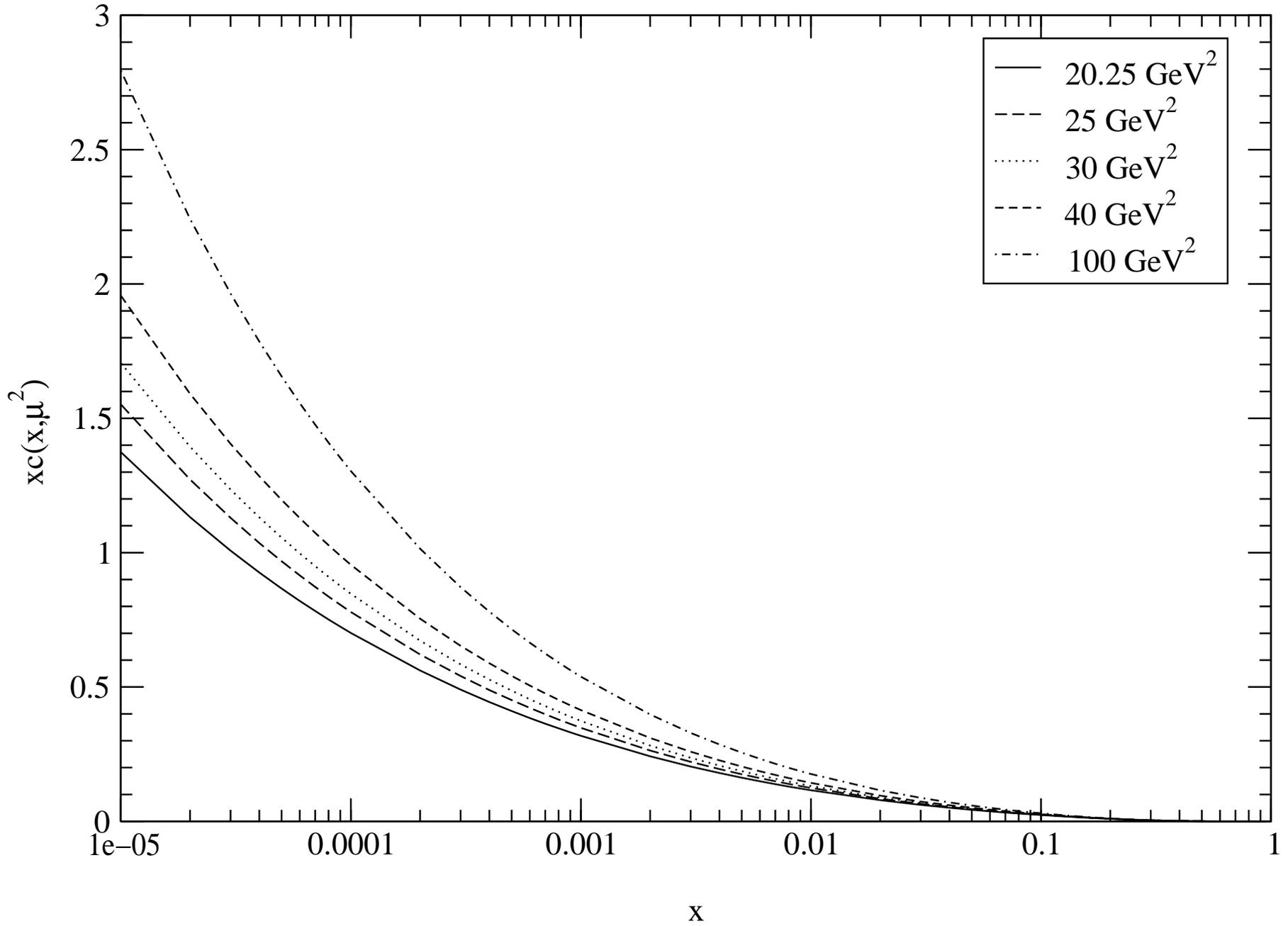
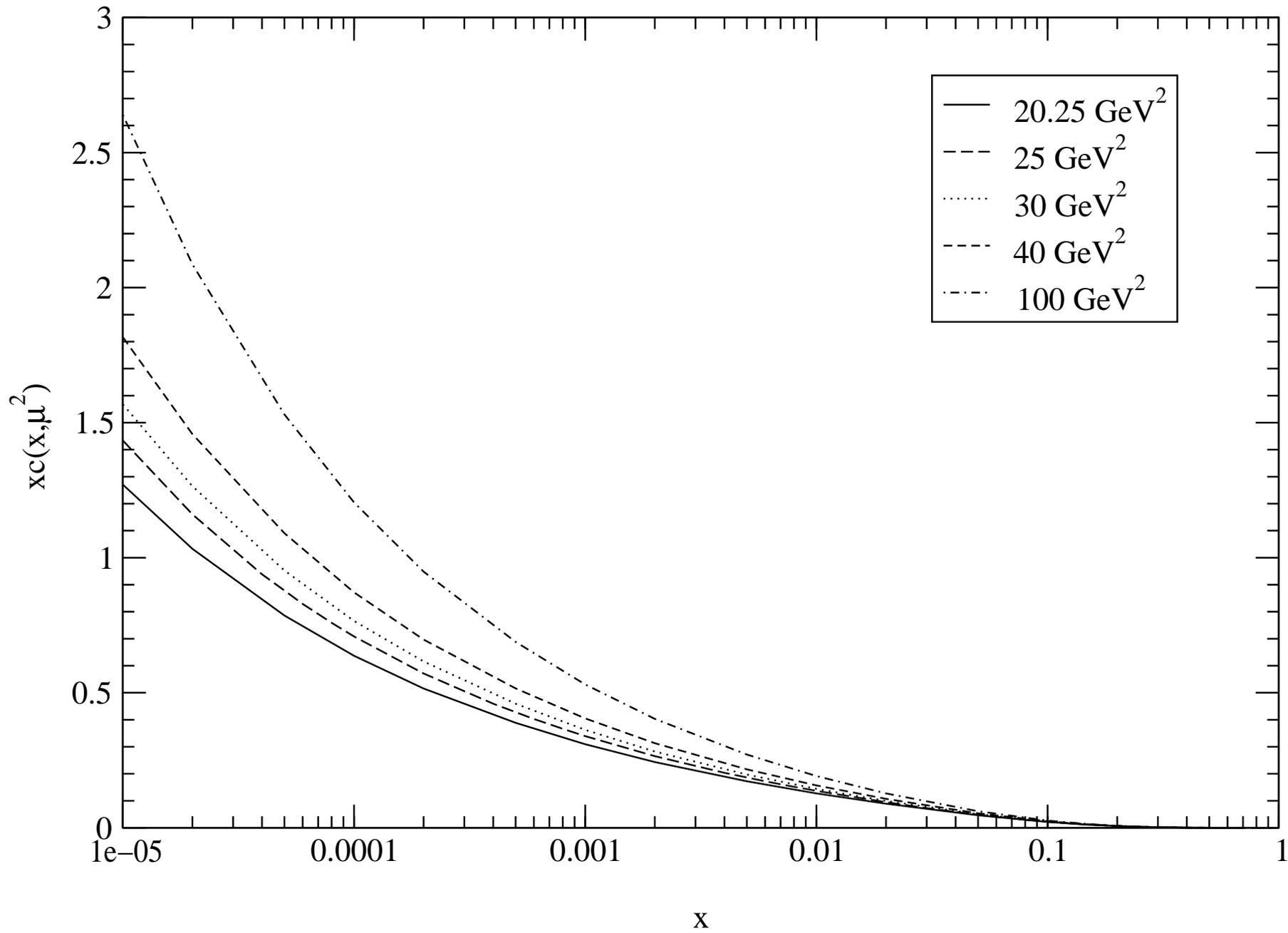
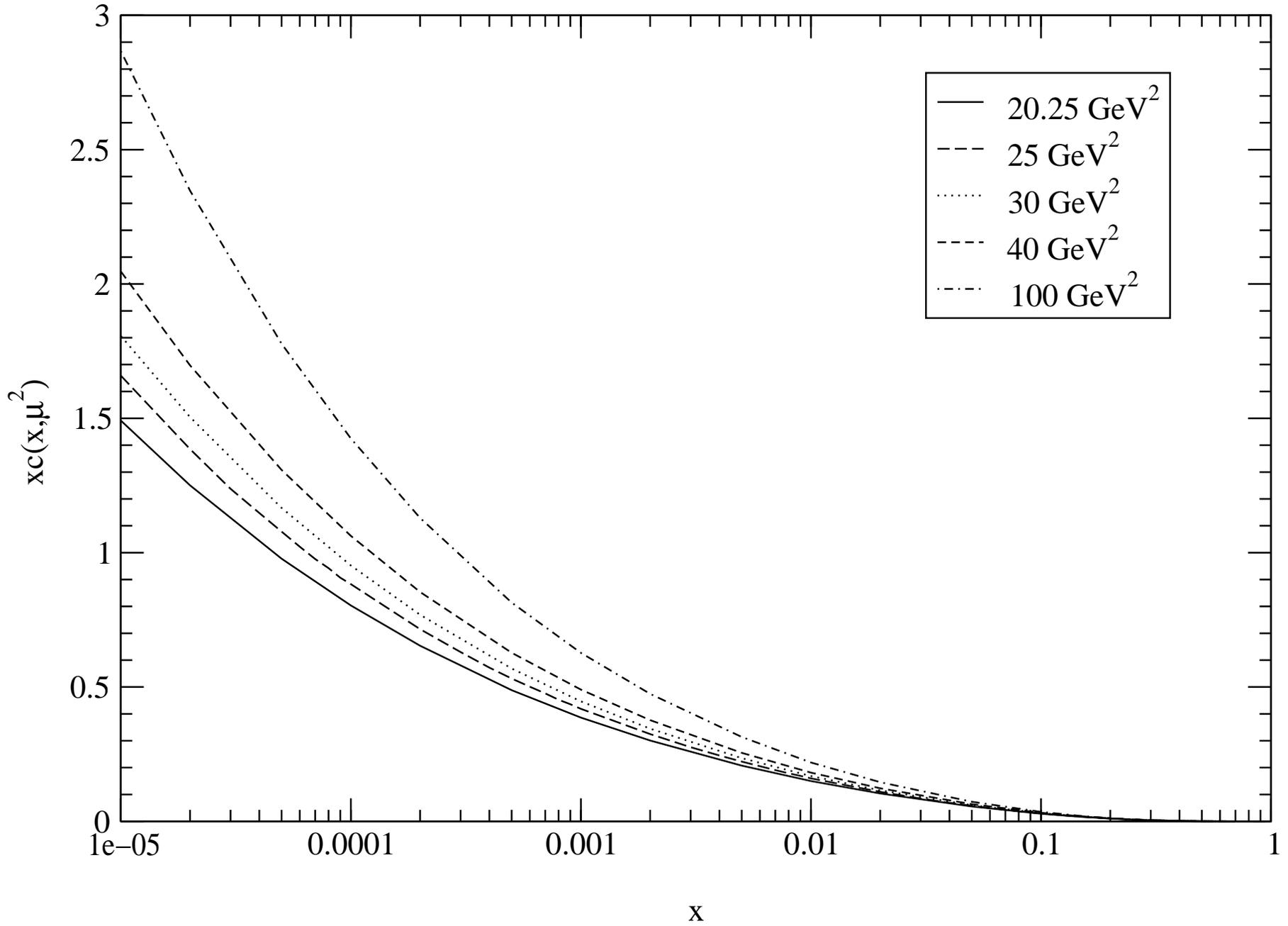
\item[Fig. 7.]
(a) The charm quark density $xc_{\rm NNLO}(5,x,\mu^2)$ in the range
$10^{-5} < x < 1$ for 
$\mu^2 =$ 20.25, 25, 30, 40 and 100 in units 
of $({\rm GeV/c}^2)^2$, 
(b) and (c) the NLO results from MRST98 set 1 and CTEQ5HQ respectively.
%--------------------------------
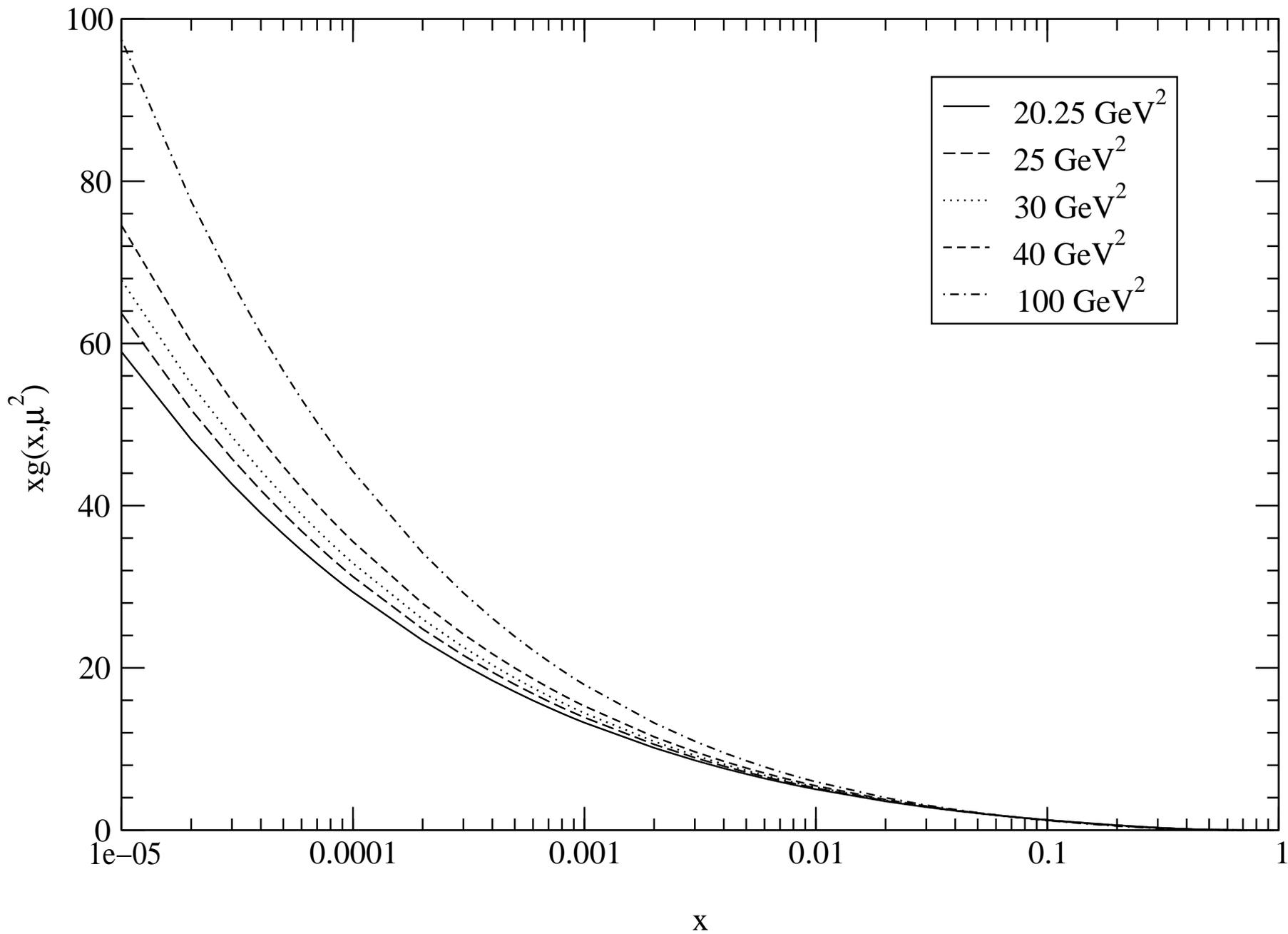
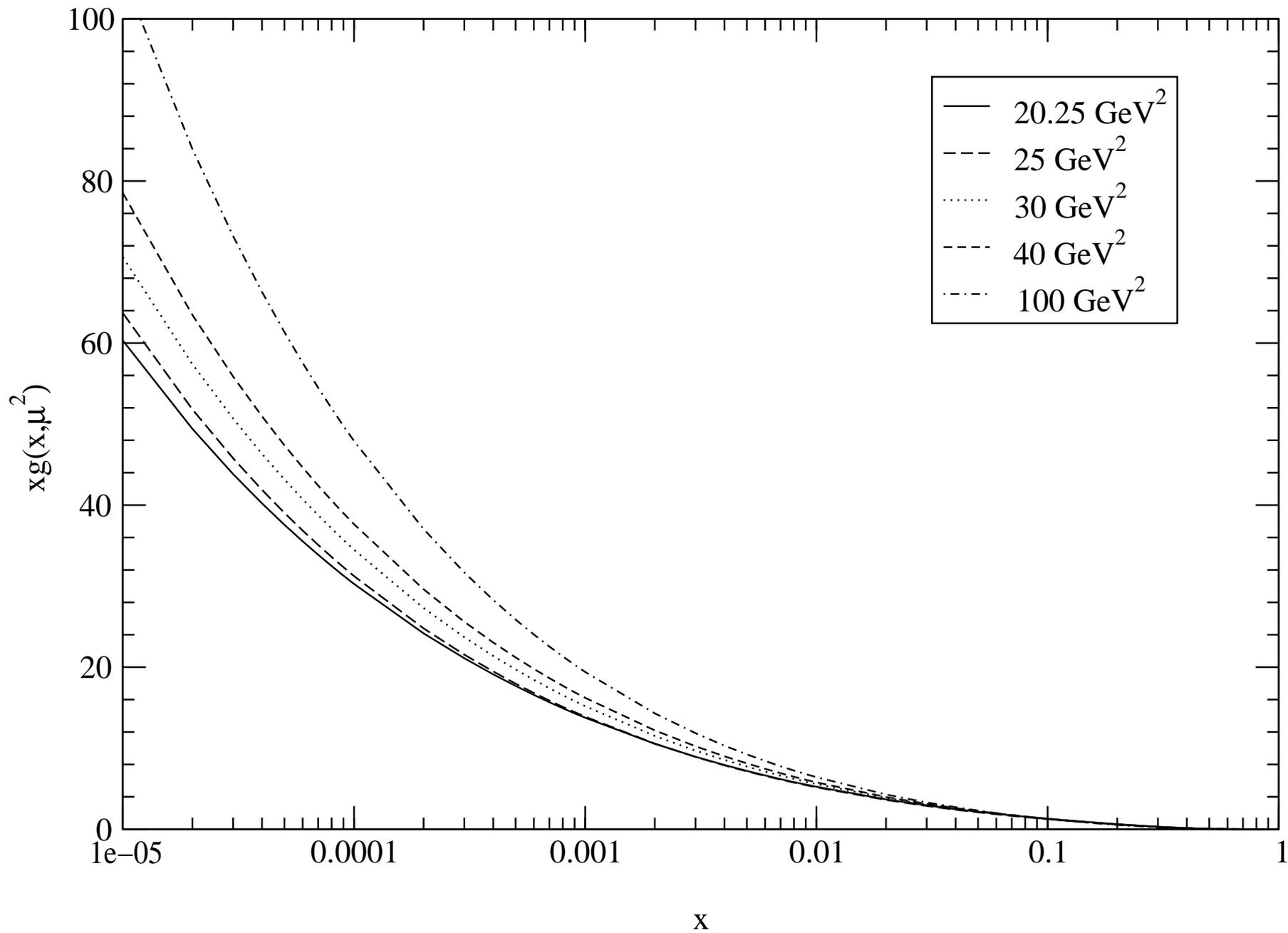
\item[Fig. 8.]
(a) The gluon density $xg_{\rm NNLO}(5,x,\mu^2)$ in the range
$10^{-5} < x < 1$ for $\mu^2 =$ 20.25, 25, 30, 40 and 100 in units 
of $({\rm GeV/c}^2)^2$, 
(b) the three-flavor NLO density.
%--------------------------------------------------
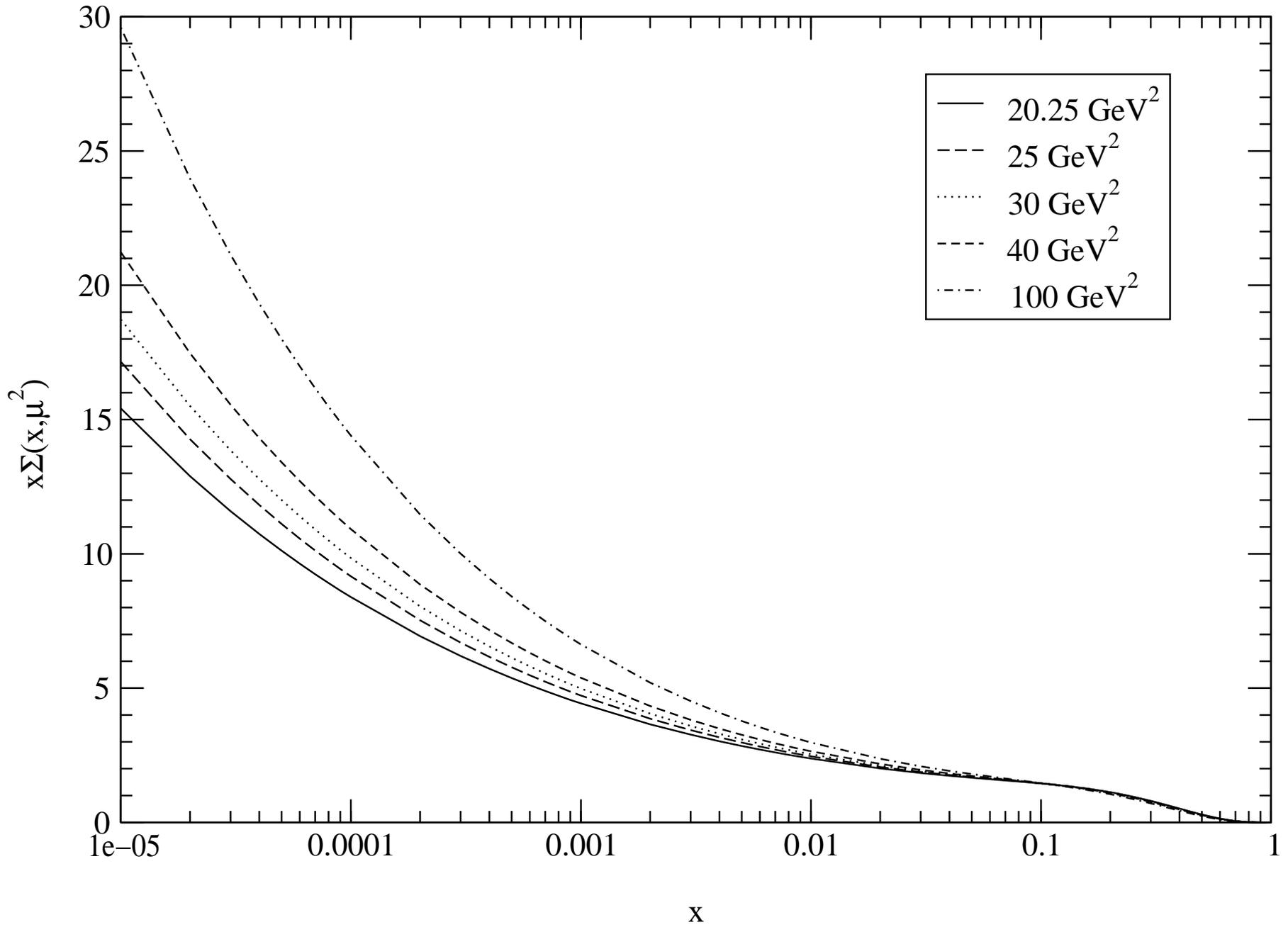
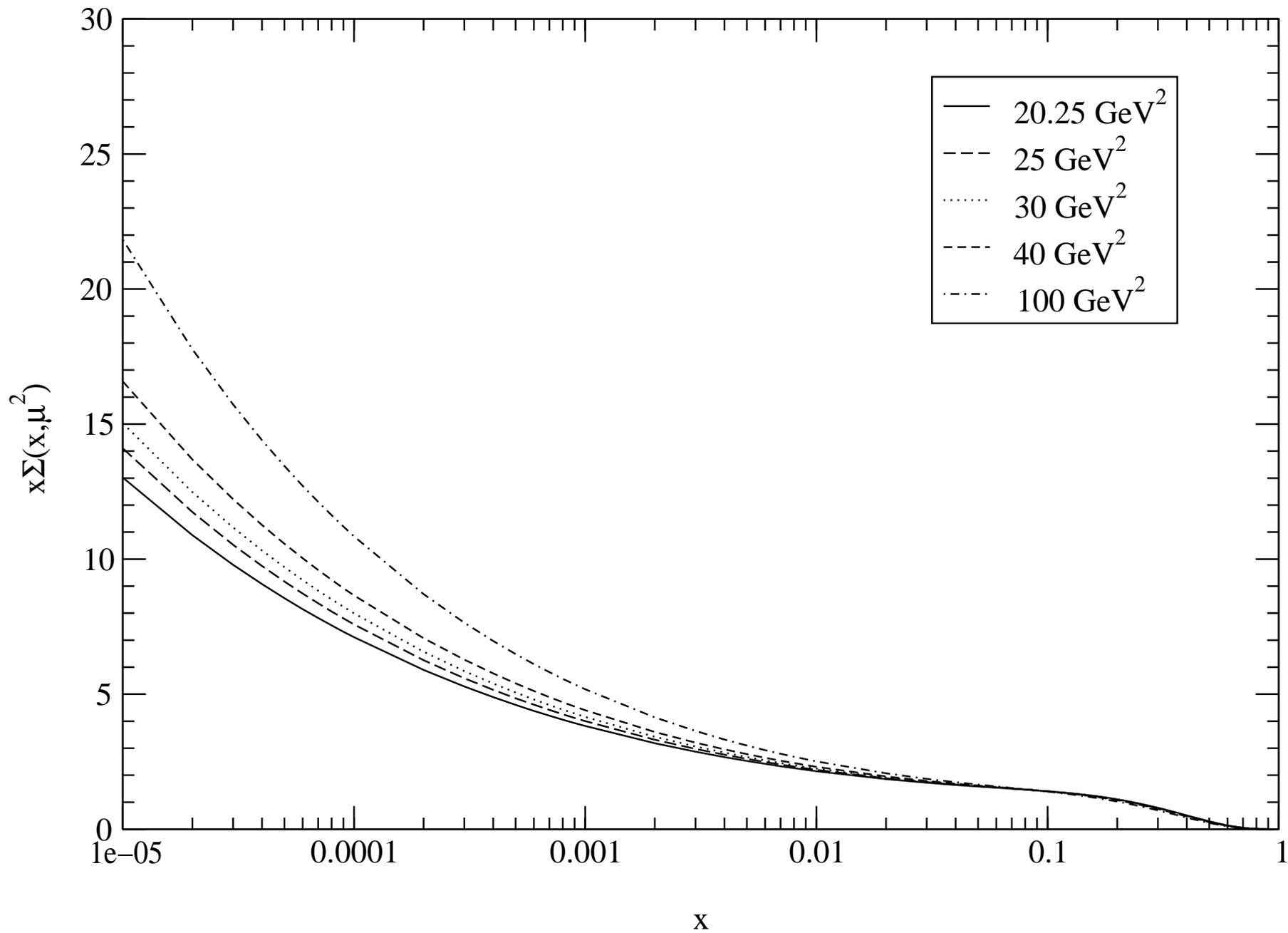
\item[Fig. 9.]
(a) The singlet quark density $x\Sigma_{\rm NNLO}(5,x,\mu^2)$ 
in the range $10^{-5} < x < 1$ for 
$\mu^2 =$ 20.25, 25, 30, 40 and 100 in units 
of $({\rm GeV/c}^2)^2$,
(b) the three-flavor NLO density. 
%------------------------------------------------------
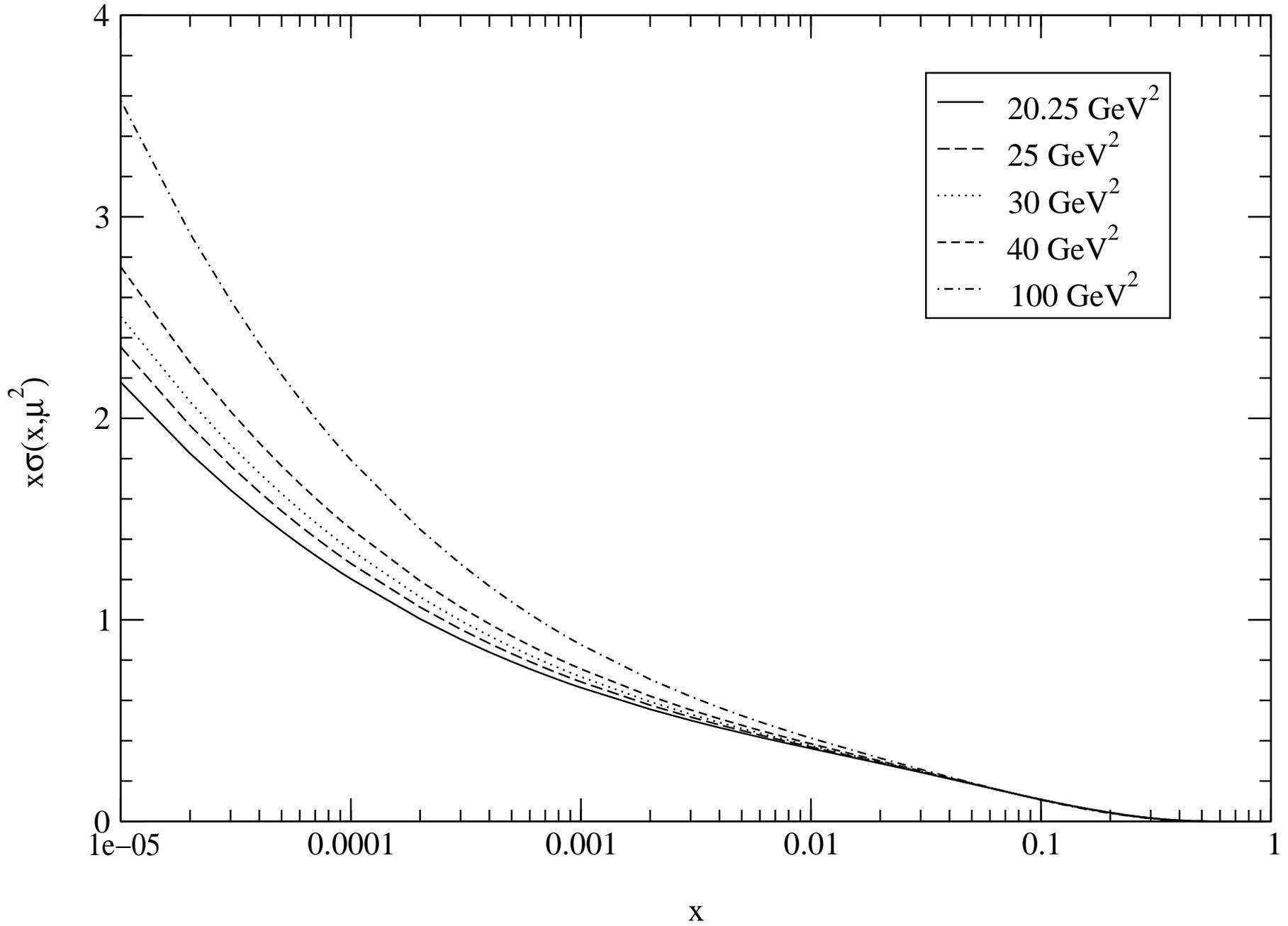
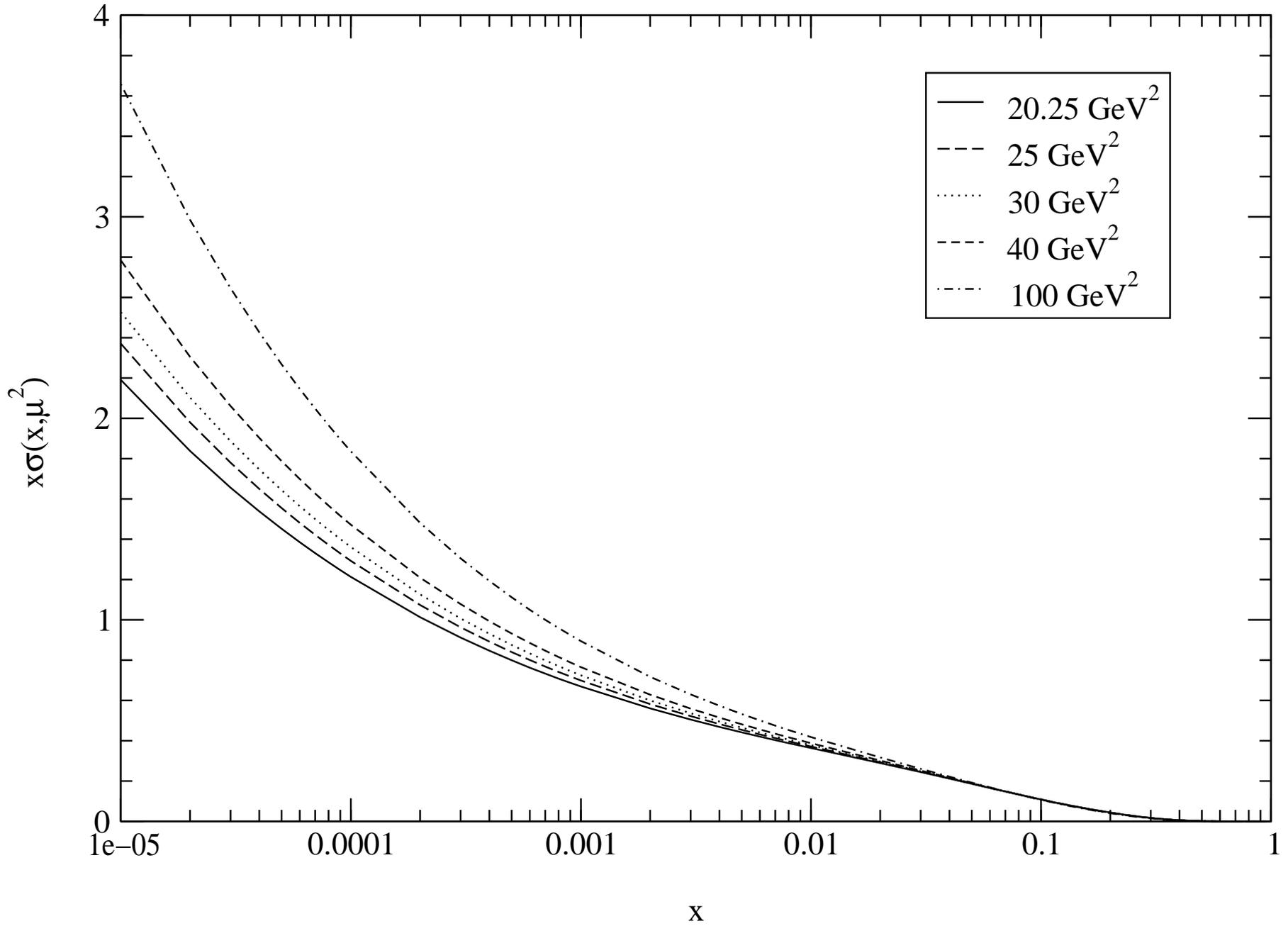
\item[Fig. 10.]
(a) The nonsinglet density $x\sigma_{\rm NNLO}(5,x,\mu^2)$ 
in the range $10^{-5} < x < 1$ for 
$\mu^2 =$ 20.25, 25, 30, 40 and 100 in units 
of $({\rm GeV/c}^2)^2$, 
(b) the three-flavor NLO density.
%-----------------------------------------------
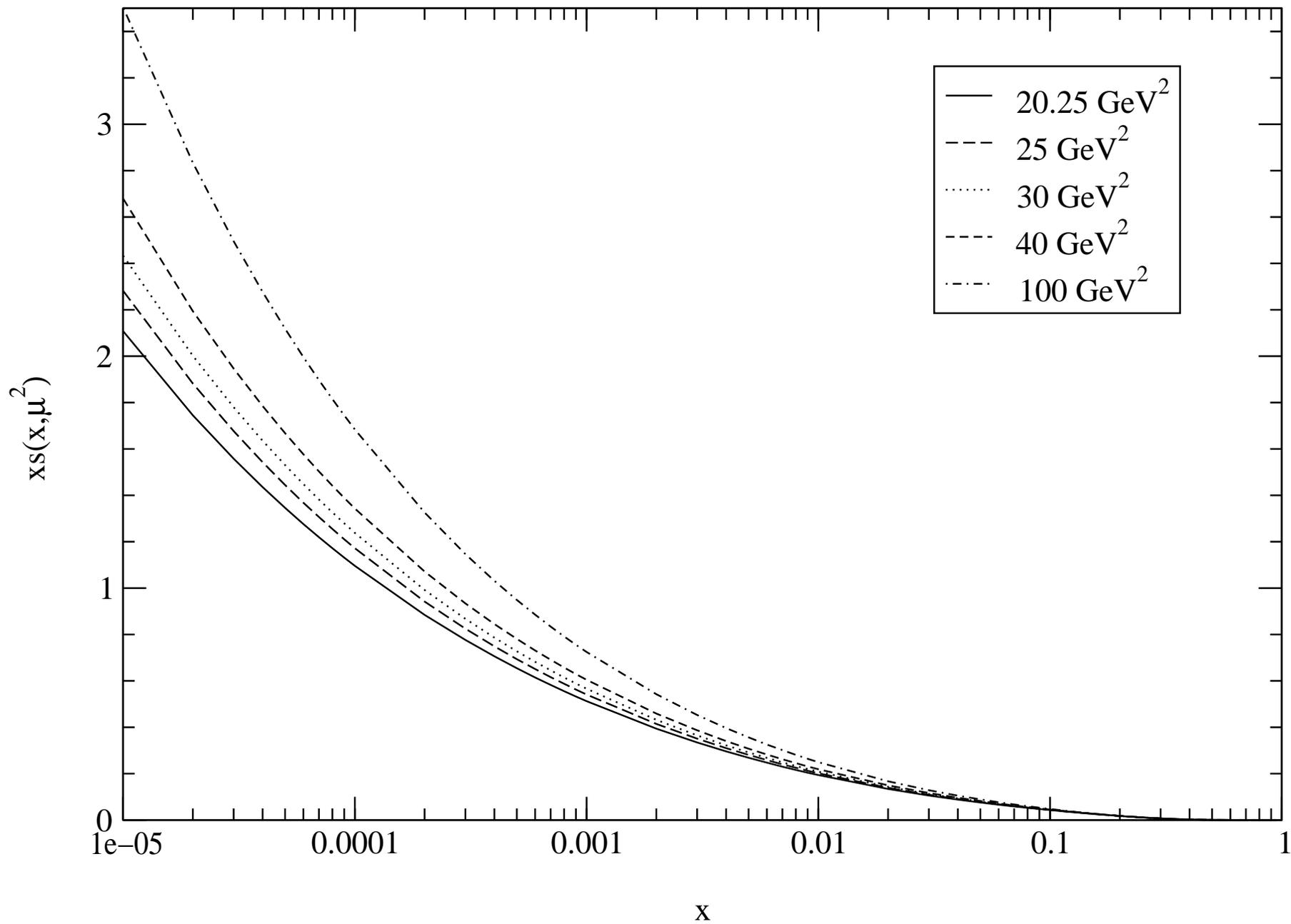
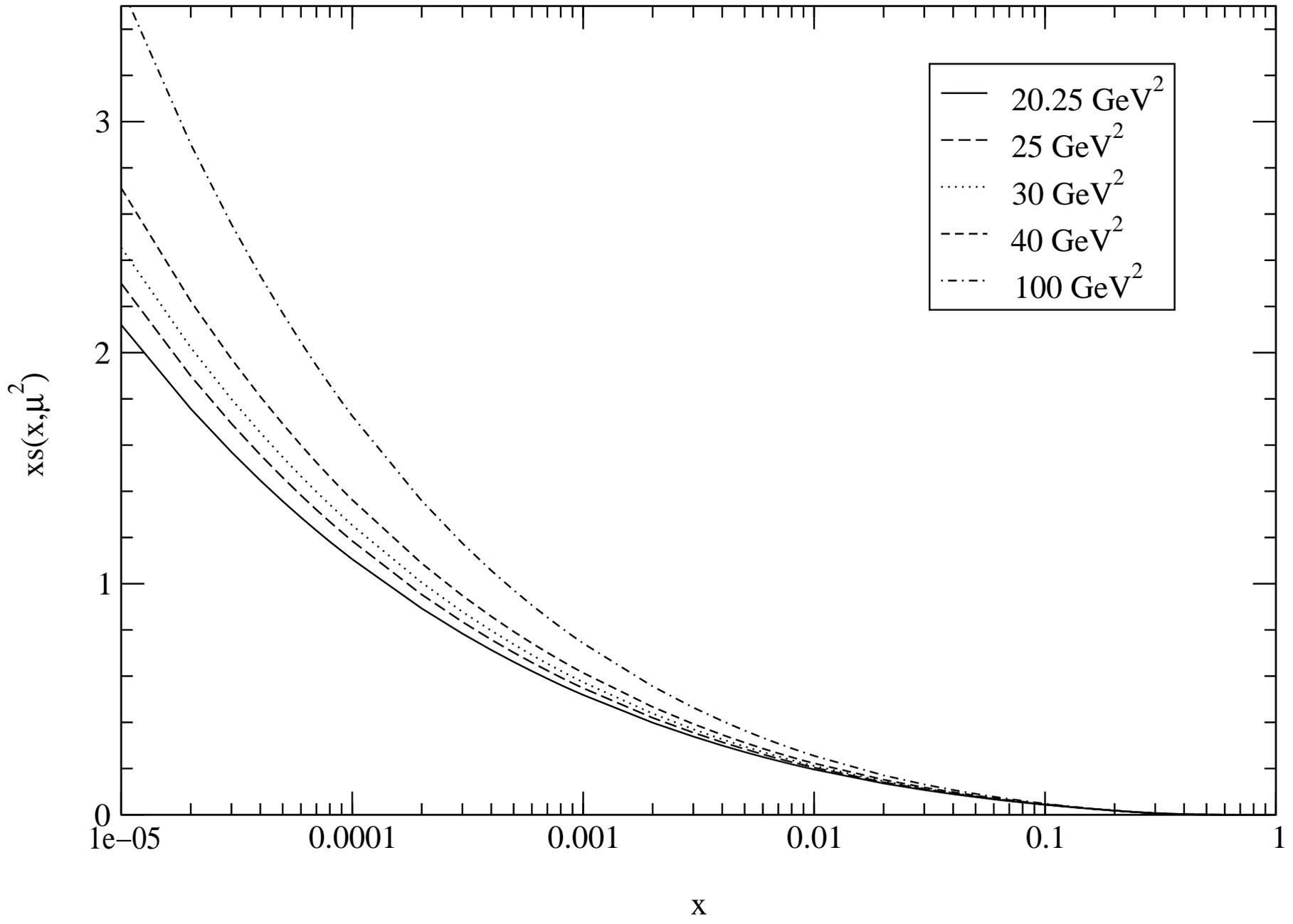
\item[Fig. 11.]
(a) The strange quark density $xs_{\rm NNLO}(5,x,\mu^2)$ 
in the range $10^{-5} < x < 1$ for $\mu^2 =$ 20.25, 30, 40 and 100 in units 
of $({\rm GeV/c}^2)^2$, 
(b) the three-flavor NLO density.
%-------------------------------------------------
\end{description}
%\end{document}

%------------------------------

\begin{thebibliography}{99}
%
%Ref(1)
\bibitem{cteq5}
H.L. Lai, J. Huston, S. Kuhlmann, J. Morf\'{\i}n, F. Olness, J. Owens,
J. Pumplin and W.K. Tung, hep-ph/9903282.
%Ref(2)
\bibitem{grv98}   M. Gl\"uck, E. Reya and A. Vogt,
Eur. Phys. J.  {\bf C5}, 461 (1998).
%Ref(3)
\bibitem{mrst98}
A.D. Martin, R.G. Roberts, W.J. Stirling and R. Thorne, 
Eur. Phys. J. {\bf C4}, 463 (1998).
%Ref(4)
\bibitem{ap}
G. Altarelli and G. Parisi, Nucl. Phys. {\bf B126}, 298 (1977).
% V.N. Gribov and L.N.Lipatov, Sov. J. Nucl. Phys. {\bf 15} 438, 675, (1972);
%Yu. Dokshitser, Sov. Phys. JETP {\bf 46}, 641 (1977).
%Ref(5)
%\bibitem{gp}
%H. Georgi and H.D. Politzer, Phys. Rev. D{\bf 9}, 416 (1974).
%Ref(6)
%\bibitem{gw}
%D.J. Gross and F. Wilczek, Phys. Rev D{\bf 9},980 (1974).
%Ref(7)
%\bibitem{frs}
%E. G. Floratos, D.A. Ross and C.T. Sachrajda, Nucl. Phys. {\bf B129}, 66 (1977)
%Erratum {\bf B139}, 545 (1978); ibid {\bf B152}, 493 (1979).
%Ref(8)
%\bibitem{gry}
%A. Gonzales-Arroyo, C. Lopez and F.J. Yndurain, Nucl. Phys. {\bf B153},
%161 (1979); 
%A. Gonzales-Arroyo and C. Lopez, Nucl. Phys. {\bf B166}, 429 (1980).
%Ref(9)
\bibitem{cfr}
G. Curci, W. Furmanski and R. Petronzio, Nucl. Phys. {\bf B175}, 27 (1980);
W. Furmanski and R. Petronzio, Phys. Lett. {\bf B97} 437, (1980);
ibid. Z. Phys. {\bf C11}, 293 (1982); the relevant NLO formulae
are presented in a convenient form in 
R.K. Ellis, W.J. Stirling and B.R. Webber, in {\it QCD and Collider Physics,}
Cambridge University Press (1996), Chapter 4.3.
%Ref(10)
%\bibitem{fkl}
%E.G. Floratos, C. Kounnas and R. Lacaze, 
%Phys. Lett. {\bf B98}, 89, 285, (1981);
%ibid. Nucl. Phys. {\bf B192}, 417 (1981).
%Ref(11)
\bibitem{mom}
S.A. Larin, T. van Ritbergen and J.A.M. Vermaseren,
Nucl. Phys. {\bf B427}, 41 (1994);
S.A. Larin et al., Nucl. Phys. {\bf B492}, 338 (1997).
%Ref(12)
\bibitem{nevo}
W.L. van Neerven and A. Vogt, hep-ph/9907472.
%Ref(13)
\bibitem{grac}
J.F. Bennett and J.A Gracey, Nucl. Phys. {\bf B417}, 241 (1998);
J.A. Gracey, Phys. Lett. {\bf B322}, 141 (1994).
%Ref(14) 
\bibitem{H1}
C. Adloff et al. (H1-collaboration), Nucl. Phys. {\bf B545}, 21 (1999).
%Ref(15)
\bibitem{ZEUS}
J. Breitweg et al. (ZEUS Collaboration), Phys. Lett. {\bf B407}, 402 (1997),
hep-ex/9908012.
%Ref(16)
\bibitem{acot}
M.A.G. Aivazis, J.C. Collins, F.I. Olness and W.-K. Tung,
Phys. Rev. D{\bf 50}, 3102 (1994); 
J.C. Collins, Phys. Rev. D{\bf 58}, 0940002 (1998).
%Ref(17)
\bibitem{thro}
R.S. Thorne and R.G. Roberts, Phys. Lett. {\bf B421}, 303 (1998);
Phys. Rev. D{\bf 57}, 6871 (1998).
%Ref(18)
\bibitem{grs} 
M. Gl\"uck, E. Reya and M. Stratmann, Nucl. Phys. {\bf B422}, 37 (1994);\\
A. Vogt in {\it Deep Inelastic Scattering and Related 
Phenomena, DIS96}, edited by G.D. 'Agostini and A. Nigro, 
(World Scientific 1997), p. 254, hep-ph/9601352.
%Ref(19)
\bibitem{lrsn} 
E. Laenen, S. Riemersma, J. Smith and W.L. van Neerven,
Nucl. Phys. {\bf B392}, 162 (1993); ibid. 229 (1993); 
S. Riemersma, J. Smith and W.L. van Neerven, Phys. Lett. {\bf B347}, 
43 (1995);
B.W. Harris and J. Smith, Nucl. Phys. {\bf B452}, 109 (1995).
%Ref(20)
\bibitem{bmsn1}
M. Buza, Y. Matiounine, J. Smith and W.L. van Neerven,
Eur. Phys. J. {\bf C1}, 301 (1998).
%Ref(21)
\bibitem{csn}
A. Chuvakin, J. Smith and W. van Neerven, hep-ph/9910250,
YITP-SB-99-15, INLO-PUB-12/99, submitted to Phys. Rev. D.
%Ref(22)
\bibitem{botje}
M. Botje, QCDNUM16: A fast QCD evolution program,
ZEUS Note 97-066.
%Ref(23)
\bibitem{brnv}
J. Bl\"umlein, S. Riemersma, W.L. van Neerven and A. Vogt, Nucl. Phys.
{\bf B}(Proc. Suppl.) {\bf 51C}, 96 (1996);\\
J. Bl\"umlein et al., 
in {\it Proceedings of the Workshop on Future Physics at HERA}
edited by G. Ingelman, A. De Roeck and R. Klanner, 
Hamburg, Germany, 25-26 Sep. 1995, p. 23, DESY 96-199, hep-ph/9609400.
%Ref(24)
\bibitem{2loop}
W. Bernreuther and W. Wetzel, Nucl. Phys. {\bf B197}, 228 (1982);
W. Bernreuther, Annals of Physics, {\bf 151}, 127 (1983).
%Ref(25)
\bibitem{2loop2}
S.A. Larin, T. van Ritbergen and J.A.M. Vermaseren, 
Nucl. Phys. {\bf B438}, 278 (1995); see also K.G. Chetyrkin,
B.A. Kniehl and M. Steinhauser, Phys. Rev. Lett. {\bf 79}, 2184 (1997).
%Ref(26)
\bibitem{miya}
M. Miyama and S. Kumano, Comput. Phys. Commun. {\bf 94}, 185 (1996);
M. Hirai, S. Kumano and M. Miyama, Comput. Phys. Commun. {\bf 108}, 38 (1998).
%Ref(27)
\bibitem{cori}
C. Coriano and S. Savkli, Comput. Phys. Commun. {\bf 118}, 236 (1999).
%Ref(28)98033398
\bibitem{riemer}
S. Riemersma, unpublished.
%Ref(27)
\bibitem{pasc}
C. Pascaud and F. Zomer, H1 Note H1-11/94-404;
V. Barone, C. Pascaud and F. Zomer, hep-ph/9907512.
%Ref(28)
%
%
\end{thebibliography}
\end{document}